\documentclass[a4paper,onecolumn,11pt,accepted=2025-04-01]{quantumarticle}
\pdfoutput=1
\usepackage[utf8]{inputenc}
\usepackage{amsmath, amssymb, amsthm}
\usepackage{bm}
\usepackage{mathtools}
\usepackage{enumerate}
\usepackage[colorlinks]{hyperref}
\usepackage[version=4]{mhchem}
\usepackage{complexity}
\usepackage[small,bf]{caption}
\usepackage{subcaption}
\usepackage[english]{babel}
\usepackage[T1]{fontenc}
\usepackage[numbers,sort,compress]{natbib}
\usepackage{color}
\usepackage{float}
\usepackage{tikz}
\usetikzlibrary{quantikz}

\DeclareMathOperator{\tr}{\mathrm{Tr}}

\theoremstyle{plain}
\newtheorem{theorem}{Theorem}

\theoremstyle{definition}
\newtheorem{definition}[theorem]{Definition}
\newtheorem{algorithm}{Algorithm}

\theoremstyle{remark}
\newtheorem{conjecture}{Conjecture}
\newtheorem{remark}[conjecture]{Remark}

\newtheorem{problem}{Problem}

\begin{document}
\title{Averaging gate approximation error and performance of Unitary Coupled Cluster ansatz in Pre-FTQC Era}
\author{Kohdai Kuroiwa}
\email{kkuroiwa@uwaterloo.ca}
\affiliation{Institute for Quantum Computing, University of Waterloo, Ontario, Canada, N2L 3G1}
\affiliation{Department of Combinatorics and Optimization, University of Waterloo}
\affiliation{Perimeter Institute for Theoretical Physics, Ontario, Canada, N2L 2Y5}
\affiliation{QunaSys Inc., Aqua Hakusan Building 9F, 1-13-7 Hakusan, Bunkyo, Tokyo 113-0001, Japan}

\author{Yuya O. Nakagawa}
\email{nakagawa@qunasys.com}
\affiliation{QunaSys Inc., Aqua Hakusan Building 9F, 1-13-7 Hakusan, Bunkyo, Tokyo 113-0001, Japan}

\maketitle

\begin{abstract}
        Fault-tolerant quantum computation (FTQC) is essential 
        to implement quantum algorithms in a noise-resilient way, 
        and thus to enjoy advantages of quantum computers even with presence of noise.
        In FTQC, a quantum circuit is decomposed into universal gates that can be fault-tolerantly implemented, for example, Clifford+$T$ gates. 
        Here, $T$ gate is usually regarded as an essential resource for quantum computation because its action cannot be simulated efficiently on classical computers and it is experimentally difficult to implement fault-tolerantly. 
        Practically, it is highly likely that only a limited number of $T$ gates are available in the near future. 
        In this \textit{Pre-FTQC} era, due to the constraint on available resources, it is vital to precisely estimate the decomposition error of a whole circuit. 
        In this paper, we propose that the Clifford+$T$ decomposition error for a given quantum circuit containing a large number of quantum gates can be modeled as the depolarizing noise by averaging the decomposition error for each quantum gate in the circuit, and our model provides more accurate error estimation than the naive estimation.
        We exemplify this by taking unitary coupled-cluster (UCC) ansatz used in the applications of quantum computers to quantum chemistry as an example.
        We theoretically evaluate the approximation error of UCC ansatz when decomposed into Clifford+$T$ gates, and the numerical simulation for a wide variety of molecules verified that our model well explains the total decomposition error of the ansatz.
        Our results enable the precise and efficient usage of quantum resources in the early-stage applications of quantum computers and fuel further research towards what quantum computation can achieve in the upcoming future. 

\end{abstract}

\section{Introduction}~\label{sec:intro}
Quantum computers are expected to exhibit superior performance over classical computers. 
For example, Shor's factoring algorithm~\cite{shor1999polynomial} shows an exponential speed-up over any known classical algorithm, and Grover's searching algorithm~\cite{grover1996fast} achieves a quadratic speed-up. 
In recent years, owing to advancements in technology, the experimental realization of quantum computers has made steady progress, and quantum supremacy has even been demonstrated using these devices~\cite{Arute2019,Zhong2020,Zhong2021,Wu2021,Daley2022,Madsen2022}. 
Noisy intermediate-scale quantum (NISQ) devices~\cite{Preskill2018} are currently largely investigated as a promising pioneer of quantum computational technologies. 
However, in the future, we aim to perform complex quantum algorithms, which show substantial quantum speed-ups, such as the phase estimation algorithm~\cite{Kitaev1995,Cleve1998} and factoring algorithm~\cite{shor1999polynomial}. 
It is inevitable that circuits required in these algorithms become highly deep, and noise occurring during the computation has significant effects without error correction. 
Thus, eventually, we need fault-tolerant quantum computation (FTQC) using quantum error correction to maximize the benefit of quantum computers.

Fault-tolerant implementation of quantum circuits with a family of error correction codes~\cite{Gottesman2009,nielsen_chuang_2010} requires an approximation of a given quantum circuit using universal gates that can be fault-tolerantly implemented with the error correction codes. 
Solovay-Kitaev (SK) theorem~\cite{Kitaev1997,Dawson2005} gives an algorithm that generates decomposition of a given quantum gate into a desired universal gate set that is closed under inversion. 
For example, SK theorem is applicable to the Clifford+$T$ universal gate set, and this set is often considered as a gate set for FTQC due to its fault-tolerant implementability ~\cite{Shor1996,Steane1997,Knill2005,Gottesman1998b,Gottesman1999,Zhou2000}. 
Hence, in this work, we focus on the circuit approximation with the Clifford+$T$ universal gate set. 
In this context, $T$ gate is considered as a crucial resource in FTQC because its fault-tolerant implementation, associating the magic state distillation~\cite{Bravyi2005}, is the most costly part in terms of both time and space overhead with current practical error-correcting codes. 
Let us also note that Gottesman-Knill Theorem~\cite{Gottesman1998a,Aaronson2004} guarantees that quantum circuits consisting only of Clifford gates are efficiently simulatable by classical computers. 

Practically, it is highly likely that we do not have a sufficient resource, \textit{e.g.,} $T$ gates and clean qubits, in the near future due to its difficulty of experimental implementation. 
In such an early stage of FTQC, \textit{Pre-FTQC era}, the precise estimation of the error arising from the decomposition of a given quantum circuit into Clifford+$T$ gates must be crucial to taking full advantage of available quantum computers.
There has been growing attention to Pre-FTQC (or early-FTQC) era recently, 
including algorithms with less auxiliary qubits~\cite{Lin2022} and improved noise robustness~\cite{Kshirsagar2024},
error reduction using error mitigation methods~\cite{Suzuki2022},  techniques shortening the circuit depth~\cite{Ding2023}, 
a method for constructing a state preparation circuit~\cite{morisaki2023classical}, 
and variational algorithms with limited number of $T$ gates~\cite{sayginel2023faulttolerant}. 
Recently, a model for this early stage of FTQC was developed by Ref.~\cite{katabarwa2023early}. In Ref.~\cite{akahoshi2023partially}, an architecture for Pre-FTQC, where we partially perform error correction, was also proposed.

Nevertheless, precise error estimation of the Clifford+$T$ decomposition of quantum circuits has not yet been thoroughly investigated.
Typically, the accuracy of such decompositions based on the SK theorem is assessed by summing the errors of individual gate decompositions via the triangle inequality~\cite{nielsen_chuang_2010}.
However, this naive evaluation may not provide a tight bound when the quantum circuit is large and the decomposition errors are non-uniform across gates.
Practical resource estimation in Pre-FTQC era may well demand more precise error estimation, which requires another way to evaluate the error. 
The best accuracy we can achieve using the available number of $T$ gates reveals how many $T$ gates will be necessary to realize desired computation, so it will greatly help the hardware development of quantum computers as well as the practical use of them.

\section{Contribution}
In this work, we propose a method for accurate estimation of the error caused by the Clifford+$T$ decomposition for a given quantum circuit, when the circuit contains a sufficiently large number of gates.  
We argue that the Clifford+$T$ approximation error of a quantum circuit can be treated as a single-qubit depolarizing channel by averaging the approximation error of each gate in the circuit, which largely simplifies the analysis of the approximation error of the whole circuit.
While the naive estimation by the triangle inequality yields the total circuit approximation error scaling linearly in given gate approximation error $\epsilon$, the error estimation given by our averaging method results is in a quadratic form of it, $O(\epsilon^2)$, as shown in Eq.~\eqref{eq:epsilon_quadratic} in Sec.~\ref{sec:average_SK}. 
In this sense, our method succeeds in reducing the estimated error of Clifford+$T$ circuit approximation. 

We illustrate our method in Unitary Coupled Cluster (UCC) ansatz~\cite{Anand2022}, which has been recognized as a good parametrized state for quantum chemistry problems in near-term quantum computing. 
This ansatz can also be applied in the Pre-FTQC era, for example, for preparation of an initial state of quantum phase estimation~\cite{Kitaev1995,Cleve1998}, so evaluation of the performance of the UCC ansatz within our framework of Pre-FTQC should be practically beneficial. 
We analyze Clifford+$T$ approximation error of the UCC ansatz using the averaging method discussed above, and show that Clifford+$T$ approximation error in this case can be expressed by a simple formula. 
Furthermore, we perform numerical simulation to verify this analysis for a wide variety of molecules. 
To see relations between the number of $T$ gates and the precision of the circuit decomposition, we develop a Clifford+$T$-gate decomposition algorithm for single-qubit rotation gates with the fixed maximum number of $T$ gates by slightly modifying the algorithm given by Ross and Selinger~\cite{Ross2016}. 
With the modified decomposition algorithm, we see that our averaging method yields precise estimation of the circuit approximation error when a given UCC ansatz is large. 
We lastly investigate how many $T$-gates are needed to realize sufficiently accurate UCC ansatz and predict the number of $T$ gates for a classically intractable large molecule.

Thus, we open up a practical perspective to implement a quantum circuit in the Pre-FTQC era. 
Our analysis and estimation will be useful to assess how many $T$ gates are needed to realize a given target quantum state in the Pre-FTQC era.
In this sense, we newly pave a way to develop an application of Pre-FTQC from a practical point of view, in which the number of available $T$-gates is restricted. 

Now we briefly compare our results to related work.
First, Campbell~\cite{Campbell2017shorter} demonstrated that the probabilistic application of quantum gates with an error of $O(\epsilon)$, sampled from an appropriate distribution, can approximate a target unitary with an improved error of $O(\epsilon^2)$.
Our findings differ from this result in that we do not rely on probabilistic implementations. Instead, the sequence of quantum gates used to approximate the target unitary is fixed.
In our numerical examples, the accumulation of errors for many quantum gates in the circuit may invoke a self-averaging effect, resulting in a quadratic error suppression of $O(\epsilon^2)$.
Second, several methods such as twirling~\cite{EndoQEM} or dynamical decoupling~\cite{Lorenza1999} are often introduced to mitigate quantum noise by averaging the effect of quantum (noisy) gates.
Our approach differs from these methods fundamentally in its setup. We do not explicitly exploit random unitary gates, and thus the gate sequence remains deterministic throughout.

The rest of this paper is organized as follows.  
In Sec.~\ref{sec:notation}, we summarize notation used in this paper. 
In Sec.~\ref{sec:average_SK}, we show our method of averaging unitary Clifford+$T$ approximation error into the depolarizing noise. 
In Sec.~\ref{sec:ucc}, we apply our analysis to the UCC ansatz. 
In Sec.~\ref{subsec:SK_error_UCC}, we show the concrete Clifford+$T$ approximation error estimation for the UCC ansatz using our averaging method. 
In Sec.~\ref{sec:algorithm}, we introduce our algorithm for generating the best Clifford+$T$-gate decomposition within some given number of $T$ gates, and compare this algorithm with the one by Ref.~\cite{Ross2016} on which it is based.
In Sec.~\ref{subsec:numerical_experiments}, we perform numerical experiments to verify our algorithm and error analysis using the UCC ansatz for various molecules in quantum chemistry. 
Finally, in Sec.~\ref{sec:discussion}, we summarize our results and discuss future directions. 

\section{Notation}
\label{sec:notation}
In this section, we introduce the notation in this paper. 
For any $\theta \in \mathbb{R}$ and any vector $\bm{n} = (n_x,n_y,n_z) \in \mathbb{R}^{3}$ satisfying $n_x^2 + n_y^2 + n_z^2 = 1$, we define the single-qubit rotation gate $R_{\bm{n}}(\theta)$ along axis $\bm{n}$ with angle $\theta$ as 
\begin{equation}
    \label{eq:definition_Rn}
    R_{\bm{n}}(\theta) = \exp\left(-i\frac{\theta}{2}\left(n_x\sigma_x + n_y\sigma_y + n_z\sigma_z \right) \right), 
\end{equation}
where $\sigma_x$, $\sigma_y$, and $\sigma_z$ are the Pauli operators defined as 
\begin{equation}
    \label{eq:definition_Pauli}
    \sigma_x = 
    \begin{pmatrix}
        0 & 1\\
        1 & 0
    \end{pmatrix}, \,\,
    \sigma_y = 
    \begin{pmatrix}
        0 & -i\\
        i & 0
    \end{pmatrix}, \,\,
    \sigma_z = 
    \begin{pmatrix}
        1 & 0\\
        0 & -1
    \end{pmatrix}
\end{equation}
in the computational basis. 
Using Eq.~\eqref{eq:definition_Pauli}, we can rewrite $R_{\bm{n}}(\theta)$ in Eq.~\eqref{eq:definition_Rn} as 
\begin{equation*}
    R_{\bm{n}}(\theta) = 
    \begin{pmatrix}
        \cos\frac{\theta}{2} - in_z\sin\frac{\theta}{2} & -i\left( n_x - in_y\right)\sin\frac{\theta}{2}\\
        -i\left( n_x + in_y\right)\sin\frac{\theta}{2} & \cos\frac{\theta}{2} + in_z\sin\frac{\theta}{2}
    \end{pmatrix}
\end{equation*}
in the computational basis. 
In particular, when $\bm{n} = (0,0,1)$, $R_{\bm{n}}(\theta)$ represents the single-qubit $z$-rotation gate with angle $\theta$, and we use $R_{z}(\theta)$ to denote this. 

For a linear operator $A$, the trace distance of $A$ is defined as 
\begin{equation*}
    \|A\|_1 \coloneqq \tr\left[\sqrt{A^\dagger A}\right], 
\end{equation*}
and the operator norm of $A$ is defined as 
\begin{equation}
\|A\| \coloneqq \max_{\ket{\phi}} \sqrt{\bra{\phi} A^\dagger A \ket{\phi}}, 
\end{equation}
where the maximum is taken over all pure states $\ket{\phi}$. 

\section{Averaging of approximation error of rotation gate  on a large parametrized circuit}~\label{sec:average_SK}
In this section, we propose an approach to analyze the effects of gate approximation error, considering a parametrized circuit that is supposed to be still used in the Pre-FTQC era. 

Typically, such a parametrized circuit, \textit{e.g.,} the UCC ansatz circuit (See Appendix~\ref{sec:AppendixA} for more details), can be expressed by Clifford gates and single-qubit $z$-rotation gates. 
Hence, when we consider Clifford+$T$ approximation of this circuit, we will consider approximation of $z$-rotation gates contained in this circuit. 
Conventionally, the total Clifford+$T$ approximation error is roughly evaluated by using the triangle inequality (See also Remark~\ref{remark:triangle} in Sec.~\ref{subsec:SK_error_UCC}), 
where Clifford+$T$ approximation error is linearly accumulated. 
However, this analysis might overestimate the total error, which would be a problem when a circuit size is large. 

Here, we propose an alternative method to evaluate the total Clifford+$T$ approximation error, which we call \textit{averaging}. 
We analyze an averaged effect of Clifford+$T$ approximation error, or the error coming from the decomposition of a $z$-rotation gate in a given parametrized circuit. 
By using this averaging analysis, we estimate the Clifford+$T$ decomposition error of the circuit. 
Following our analysis, one may model the effective behavior of Clifford+$T$ approximation error by a depolarizing channel. 
The averaged effect we derive is expected to be valid when there are a large number of gates to be decomposed so that Clifford+$T$ approximation error of each gate in the circuit will be well-averaged over the whole circuit. 

Note that while our main focus is Clifford+$T$ approximation for a parametrized quantum circuit, the averaging method we propose below applies to any decomposition of $z$-rotation. 

Now, let $R_z(\theta)$ be the desired single-qubit $z$-rotation gate with angle $\theta$, and suppose that we have a unitary gate $U$ as an approximation of $R_z(\theta)$ to accuracy $\epsilon > 0$, that is,  
\begin{equation}~\label{eq:SK_approx}
    \|U - R_z(\theta)\| \leq \epsilon. 
\end{equation}
When we try to apply the gate $R_z(\theta)$, 
we actually apply the noisy gate $U = (UR_z^\dagger(\theta)) R_z(\theta)$. Thus, $UR_z^\dagger(\theta)$ is considered to represent the unitary noise induced by the decomposition. 
Since any single-qubit unitary is equivalent to some single-qubit rotation gate $R_{\bm{n}}(\varphi)$ along axis $\bm{n} = (n_x,n_y,n_z)$ with angle $\varphi$ 
up to some global phase $\mathrm{e}^{i\alpha}$, 
we can express the unitary noise as $UR_z^\dagger(\theta) = \mathrm{e}^{i\alpha} R_{\bm{n}}(\varphi)$ using real numbers $\alpha, \varphi \in (-\pi,\pi]$ and a normalized real vector $\bm{n}$.  
Since the operator norm is unitary-invariant, from Eq.~\eqref{eq:SK_approx}, we have
\begin{equation}~\label{eq:Rn_approx}
    \|I - \mathrm{e}^{i\alpha} R_{\bm{n}}(\varphi)\| \leq \epsilon.
\end{equation}
To satisfy Eq.~\eqref{eq:Rn_approx}, all the singular values of $I - \mathrm{e}^{i\alpha} R_{\bm{n}}(\varphi)$ must be upper bounded by $\epsilon$. 
Consider the following positive matrix: 
\begin{equation*}
\begin{aligned}
    &(I - \mathrm{e}^{i\alpha} R_{\bm{n}}(\varphi))(I - \mathrm{e}^{i\alpha} R_{\bm{n}}(\varphi))^\dagger \\
    &= 2I - \mathrm{e}^{i\alpha} R_{\bm{n}}(\varphi) - \mathrm{e}^{-i\alpha} R_{\bm{n}}^\dagger(\varphi) \\ 
    &= 2 \left(
    \begin{array}{cc}
        1 - \cos\alpha \cos \tfrac{\varphi}{2} -  n_z \sin\alpha \sin \tfrac{\varphi}{2}& - (n_x-in_y)\sin\alpha \sin \tfrac{\varphi}{2}\\
        - (n_x+in_y)\sin\alpha \sin \tfrac{\varphi}{2} & 1 - \cos\alpha \cos \tfrac{\varphi}{2} +  n_z \sin\alpha \sin \tfrac{\varphi}{2}
    \end{array}
    \right).
\end{aligned}
\end{equation*}
The eigenvalues of this matrix are 
\begin{equation*}
    2\left(1 - \cos\left(\frac{\varphi}{2} \pm \alpha\right)\right). 
\end{equation*}
Therefore, for Eq.~\eqref{eq:Rn_approx} to be satisfied, the following must hold:
\begin{align}
    \label{eq:condition1}
    &1 - \cos\left(\frac{\varphi}{2} + \alpha\right) \leq \frac{\epsilon^2}{2}, \\
    \label{eq:condition2}
    &1 - \cos\left(\frac{\varphi}{2} - \alpha\right) \leq \frac{\epsilon^2}{2}.
\end{align}
Define $\delta \coloneqq \cos^{-1}(1-\epsilon^2/2)$. 
By this definition, when $\epsilon \approx 0$, we also have $\delta \approx 0$. 
Since $\alpha,\varphi \in (-\pi,\pi]$, these conditions, Eq.~\eqref{eq:condition1} and \eqref{eq:condition2}, are equivalent to 
\begin{align}
    &-2\delta-2\alpha \leq \varphi \leq 2\delta-2\alpha, \\ 
    &-2\delta+2\alpha \leq \varphi \leq 2\delta+2\alpha,
\end{align}
that is,  $\alpha$ and $\varphi$ must satisfy
\begin{equation*}
    \left|\frac{\varphi}{2} \pm \alpha\right| \leq \delta. 
\end{equation*}
Additionally, $\bm{n}$ is constrained by the condition $|\bm{n}|^2 = n_x^2 + n_y^2 + n_z^2 = 1$. 

Now, we analyze the average effect of the error caused by the approximation of a single qubit $z$-rotation. 
We show the effect of the noise $UR_z^\dagger(\theta) = \mathrm{e}^{i\alpha} R_{\bm{n}}(\varphi)$ averaged over the parameters is indeed equivalent to the depolarizing noise. 
Assuming that the effect of noise is completely random over the circuit, we take the Haar measure for the average. 
Hence, the average effect of $R_{\bm{n}}(\varphi)$ on a single qubit operator $A$ can be expressed as 
\begin{equation}
\label{eq:average_effect}
\int_{\left|\frac{\varphi}{2} \pm \alpha\right| \leq \delta} \sin^2\frac{\varphi}{2}d\varphi d\alpha\int_{|\bm{n}|=1} dn_x dn_y dn_z R_{\bm{n}}(\varphi) A R_{\bm{n}}^\dagger (\varphi), 
\end{equation}
where the term $\sin^2\frac{\varphi}{2}$ comes from the Haar measure. 
Note that the normalization is not yet taken into account at this point. 

First, we evaluate the integral with respect to $\bm{n}$ in Eq.~\eqref{eq:average_effect}. 
The integral can be written as 
\begin{align*}
    &\int_{|\bm{n}|=1} dn_x dn_y dn_z R_{\bm{n}}(\varphi) A R_{\bm{n}}^\dagger (\varphi) \\ 
    &= \int_{|\bm{n}|=1} dn_x dn_y dn_z \Bigg(\cos^2\frac{\varphi}{2}A + \sin^2\frac{\varphi}{2}\left(n_x^2 \sigma_xA\sigma_x + n_y^2 \sigma_yA\sigma_y + n_z^2 \sigma_zA\sigma_z \right)\\
    &\qquad +i\cos\frac{\varphi}{2}\sin\frac{\varphi}{2}\left( n_x(A\sigma_x - \sigma_x A) + n_y(A\sigma_y - \sigma_y A) + n_z(A\sigma_z - \sigma_z A) \right)\\ 
    &\qquad +\sin^2\frac{\varphi}{2} \left(n_xn_y(\sigma_x A \sigma_y + \sigma_y A \sigma_x) + n_yn_z(\sigma_y A \sigma_z + \sigma_z A \sigma_y) + n_zn_x(\sigma_z A \sigma_x + \sigma_x A \sigma_z) \right)
    \Bigg). 
\end{align*}
Since 
\begin{align*}
    &\int_{|\bm{n}|=1} dn_x dn_y dn_z n_i = 0 \\
    &\int_{|\bm{n}|=1} dn_x dn_y dn_z n_jn_k = 0
\end{align*}
for all $i = x,y,z$ and $j,k = x,y,z$ with $j\neq k$, 
the third and the fourth terms in the integral will be zero. 
Thus, using the fact that 
\begin{align*}
    &\int_{|\bm{n}|=1} dn_x dn_y dn_z = 4\pi, \\ 
    &\int_{|\bm{n}|=1} dn_x dn_y dn_z n_i^2 = \frac{4\pi}{3}
\end{align*}
for $i = x,y,z$, we have 
\begin{equation*}
     \int_{|\bm{n}|=1} dn_x dn_y dn_z R_{\bm{n}}(\varphi) A R_{\bm{n}}^\dagger (\varphi) = 4\pi\left(\cos^2 \frac{\varphi}{2} A + \frac{1}{3}\sin^2\frac{\varphi}{2} \left(\sigma_xA\sigma_x + \sigma_y A\sigma_y + \sigma_zA\sigma_z\right)\right). 
\end{equation*}
The normalization constant for this integral is 
\begin{equation*}
    \int_{|\bm{n}|=1} dn_x dn_y dn_z = 4\pi. 
\end{equation*}
Hence, the resulting average operator is 
\begin{equation*}
    \cos^2 \frac{\varphi}{2} A + \frac{1}{3}\sin^2\frac{\varphi}{2} \left(\sigma_xA\sigma_x + \sigma_y A\sigma_y + \sigma_zA\sigma_z\right). 
\end{equation*}
Now, we evaluate the integral with respect to $\alpha$ and $\delta$. 
Observe that 
\begin{align*}
&\int_{\left|\frac{\varphi}{2} \pm \alpha\right| \leq \delta} \sin^2\frac{\varphi}{2} \cos^2 \frac{\varphi}{2}d\varphi d\alpha = \frac{8\delta^2 + \cos 4\delta -1}{16}, \\ 
&\int_{\left|\frac{\varphi}{2} \pm \alpha\right| \leq \delta} \sin^4\frac{\varphi}{2} d\varphi d\alpha = \frac{24\delta^2 + 16\cos2\delta - \cos 4\delta -15}{16}. 
\end{align*} 
Note that the normalization constant for this integral is given by 
\begin{equation*}
    \int_{\left|\frac{\varphi}{2} \pm \alpha\right| \leq \delta} \sin^2\frac{\varphi}{2} d\varphi d\alpha = 2\delta^2 + \cos2\delta -1. 
\end{equation*}
Thus, after the normalization, the resulting operator is 
\begin{equation*}
\frac{8\delta^2 + \cos 4\delta -1}{16(2\delta^2 + \cos2\delta -1)} A + \frac{1}{3} \frac{24\delta^2 + 16\cos2\delta - \cos 4\delta -15}{16(2\delta^2 + \cos2\delta -1)} \left(\sigma_xA\sigma_x + \sigma_y A\sigma_y + \sigma_zA\sigma_z\right). 
\end{equation*}
Since we have Taylor expansions 
\begin{align*}
    &\frac{8\delta^2 + \cos 4\delta -1}{16(2\delta^2 + \cos2\delta -1)} = 1 - \frac{2\delta^2}{5} + O(\delta^4),\\ 
    &\frac{24\delta^2 + 16\cos2\delta - \cos 4\delta -15}{16(2\delta^2 + \cos2\delta -1)} = \frac{2\delta^2}{5} + O(\delta^4), 
\end{align*}
when $\delta$ is sufficiently small, the resulting average operator can approximately be written as 
\begin{equation*}
    \left(1- \frac{2\delta^2}{5}\right)A + \frac{2\delta^2}{15}\left(\sigma_xA\sigma_x + \sigma_y A\sigma_y + \sigma_zA\sigma_z\right) = \left(1- \frac{8\delta^2}{15}\right)A + \frac{8\delta^2}{15} \tr(A) \frac{I}{2}. 
\end{equation*}
In summary, the average effect on $A$ can be written as 
\begin{equation*}
    A \mapsto \left(1- \frac{8\delta^2}{15}\right)A + \frac{8\delta^2}{15} \tr(A) \frac{I}{2}; 
\end{equation*}
that is, the effect is equivalent to the single-qubit depolarizing noise with error probability
\begin{equation}
\label{eq:epsilon_quadratic}
    p \coloneqq \frac{8\delta^2}{15}  = O(\epsilon^2), 
\end{equation}
where we used the Taylor series 
\begin{equation*}
    \cos^{-1}(1-x) = \sqrt{2x} + O(x^{3/2}). 
\end{equation*}
Hence, when averaged, the unitary gate-decomposition error of size $\epsilon$ is represented as depolarizing error whose magnitude is $O(\epsilon^2)$. 
Note that the naive estimation by the triangle inequality gives $O(\epsilon)$~\cite{nielsen_chuang_2010}. We will exemplify this with the UCC ansatz in Eq.~\eqref{eq:naive_epsilon} in Sec.~\ref{subsec:SK_error_UCC}. 
Hence, we obtained a completely new error formula by averaging. 
We also justify our analysis by considering a "variance" operator with respect to averaging in Appendix~\ref{appendix:variance}. 
As noted at the beginning of this section, we did not restrict the approximation to a specific form, so the analysis conducted in this section applies to any approximation method of $z$-rotation gates. 

We believe that the averaged error derived in this section appropriately models the effects of Clifford+$T$ approximation error of a large quantum circuit, which consists of many single-qubit rotations. 
Clifford+$T$ approximation error originated from the Clifford+$T$ decomposition of each single-qubit rotation gate is absolutely unitary by definition. 
At the same time, when we look at the effects of all of these unitary errors over the large circuit, it is expected that their effects look to be averaged. 
Therefore, from our analysis here, we anticipate that Clifford+$T$ approximation error of a large quantum circuit with many single-qubit rotations is well-modeled by the single-qubit depolarizing noise. 
In Sec.~\ref{sec:ucc}, we numerically examine this using Unitary Coupled Cluster (UCC) ansatz.

\section{Application to Unitary Coupled Cluster (UCC) ansatz}~\label{sec:ucc}
In this section, we investigate Clifford+$T$ approximation error of the UCC ansatz both theoretically and numerically, as an illustration of our results. 
In Sec.~\ref{subsec:SK_error_UCC}, we analyze how the averaged Clifford+$T$ approximation error on the UCC ansatz behaves based on the theoretical argument of Sec.~\ref{sec:average_SK}. 
In Sec.~\ref{sec:algorithm}, we present our algorithm for generating the best Clifford+$T$-gate decomposition within some given number of $T$ gates, which is motivated by the algorithm given in Ref.~\cite{Ross2016}.
In Sec.~\ref{subsec:numerical_experiments}, we numerically investigate Clifford+$T$ approximation error of the UCC ansatz for various molecular systems in quantum chemistry when the number of $T$ gates is restricted. 
The averaged Clifford+$T$ approximation error well describes the numerical results and we also estimate the actual number of the $T$ gates required to represent a quantum state with sufficient accuracy for the first time.

In the Pre-FQTC era, while we cannot perform powerful quantum algorithms that require tremendous amounts of resources, we may still wish to prepare specific quantum states that are impossible to realize on near-term quantum devices without error correction,  \textit{e.g.,} parametrized quantum states realized by a relatively deep ansatz circuit.   
In quantum chemistry, although the quantum phase estimation~\cite{Kitaev1995,Cleve1998} is expected to compute the ground state energy of a given molecular Hamiltonian efficiently, its realization will require about $O(10^{10}) \sim O(10^{15})$ $T$ gates for molecules of practical interest~\cite{Reiher2017, Babbush2018, Berry2019qubitizationof, vonBurg2021, Lee2021, yoshioka2022hunting}.
It is then natural to consider using early FTQC with a smaller number of $T$-gates, such as the variational quantum eigensolver~\cite{Peruzzo2014} with an ansatz with a deep quantum circuit, at some point in time.
In the context of near-term quantum computation, the UCC ansatz~\cite{Anand2022}, which is based on the coupled cluster theory~\cite{Bartlett2007}, has been considered as a nice parametrized state (ansatz state) to represent a ground state of a molecular Hamiltonian. 
While the application of the UCC ansatz has been primarily discussed for near-term quantum devices so far, we believe that this ansatz will rather be beneficial in the Pre-FTQC era because its circuit tends to be pretty deep in general and typically beyond the capability of near-term quantum devices~\footnote{
The UCC singles and doubles ansatz, which is used in the numerical simulation below, requires $O(N^4)$ gates for $N$ molecular orbitals in a naive implementation.
}.
We also note that the UCC ansatz with some fixed parameters can serve as an initial state of quantum phase estimation, which may overcome the problem of the poor overlap between the initial state and the exact ground state of the molecular Hamiltonian~\cite{lee2022there}.
We summarize the definition and properties of the UCC ansatz in Appendix~\ref{sec:AppendixA} (see also Ref.~\cite{Anand2022} for more details). 

\subsection{Averaged effects of Clifford+$T$ approximation error on UCC ansatz}~\label{subsec:SK_error_UCC}
We evaluate Clifford+$T$ approximation error of the UCC ansatz with large number of qubits by using the average effects of Clifford+$T$ approximation error derived in Sec.~\ref{sec:average_SK}.
Observing that quantum gates in the UCC ansatz can be expressed as a composition of $H$ gates, $S$ gates, CNOT gates, and parametrized single-qubit $z$-rotation gates, (see Appendix~\ref{sec:AppendixA} for the case of UCC singles and doubles ansatz),
we know that Clifford+$T$ approximation error arises only when we approximate the $z$-rotation by the Clifford+$T$ gate decomposition. 
Suppose that we have a UCC ansatz circuit with $L$ $z$-rotation gates. 
Then, the UCC ansatz circuit with the averaged Clifford+$T$ approximation error can be modeled as a quantum operation 
\begin{equation*}
    \mathcal{N}_{\textup{UCC}} \coloneqq \mathcal{E}_L\circ\mathcal{U}_L\circ\mathcal{E}_{L-1}\circ\cdots \circ \mathcal{E}_{2}\circ\mathcal{U}_2\circ\mathcal{E}_1\circ\mathcal{U}_1,  
\end{equation*}
where $\mathcal{E}_i$ is the single-qubit depolarizing channel acting on the single-qubit system to which the $i$th $z$-rotation is applied, and $\mathcal{U}_i$ is a unitary operation composed of Clifford gates and only one $z$-rotation gate, for all $1\leq i \leq L$. 
Denoting the initial input to this ansatz $\rho_0$, the total effects of Clifford+$T$ approximation error of the circuit can be evaluated by 
\begin{equation*}
   \epsilon_{\textup{tot}} \coloneqq \|\mathcal{N}_{\textup{UCC}}(\rho_0) - \mathcal{U}_{\textup{UCC}}(\rho_0)\|_1, 
\end{equation*}
where 
\begin{equation*}
    \mathcal{U}_{\textup{UCC}} \coloneqq \mathcal{U}_L\circ\cdots \circ\mathcal{U}_2\circ\mathcal{U}_1
\end{equation*}
is the given UCC ansatz circuit and $\|\cdot\|_1$ is the trace norm.
Observe that 
\begin{equation*}
    \begin{aligned}
    &\mathcal{N}_{\textup{UCC}}(\rho_0) \\
    &= \mathcal{E}_L\circ\mathcal{U}_L\circ\mathcal{E}_{L-1}\circ\cdots \circ \mathcal{E}_{2}\circ\mathcal{U}_2\circ\mathcal{E}_1\circ\mathcal{U}_1(\rho_0) \\ 
    &= \mathcal{E}_L\circ\mathcal{U}_L\circ\mathcal{E}_{L-1}\circ\cdots \circ \mathcal{E}_{2}\circ\mathcal{U}_2\circ\left((1-p)\mathcal{U}_1(\rho_0) + p\Delta_1\circ\mathcal{U}_1(\rho_0)\right) \\ 
    &= (1-p)\mathcal{E}_L\circ\mathcal{U}_L\circ\mathcal{E}_{L-1}\circ\cdots \circ \mathcal{E}_{2}\circ\mathcal{U}_2\circ\mathcal{U}_1(\rho_0)  \\
    &\quad\quad+ p\mathcal{E}_L\circ\mathcal{U}_L\circ\mathcal{E}_{L-1}\circ\cdots \circ \mathcal{E}_{2}\circ\Delta_1\circ\mathcal{U}_1(\rho_0)\\
    &= (1-p)\mathcal{E}_L\circ\mathcal{U}_L\circ\mathcal{E}_{L-1}\circ\cdots \circ \mathcal{E}_{2}\circ\mathcal{U}_2\circ\mathcal{U}_1(\rho_0) + p\rho_1\\
    &= (1-p)\mathcal{E}_L\circ\mathcal{U}_L\circ\mathcal{E}_{L-1}\circ\cdots \circ \mathcal{E}_{3}\circ\mathcal{U}_3\left((1-p)\mathcal{U}_2\circ\mathcal{U}_1(\rho_0) + p\Delta_2\circ\mathcal{U}_2\circ\mathcal{U}_1(\rho_0)\right) + p\rho_1\\
    &= (1-p)^2\mathcal{E}_L\circ\mathcal{U}_L\circ\mathcal{E}_{L-1}\circ\cdots \circ \mathcal{E}_{3}\circ\mathcal{U}_3\circ\mathcal{U}_2\circ\mathcal{U}_1(\rho_0) \\ 
    &\quad\quad
    + p(1-p)\mathcal{E}_L\circ\mathcal{U}_L\circ\mathcal{E}_{L-1}\circ\cdots \circ \mathcal{E}_{3}\circ\Delta_2\circ\mathcal{U}_2\circ\mathcal{U}_1(\rho_0)+p\rho_1\\
    &= (1-p)^2\mathcal{E}_L\circ\mathcal{U}_L\circ\mathcal{E}_{L-1}\circ\cdots \circ \mathcal{E}_{3}\circ\mathcal{U}_3\circ\mathcal{U}_2\circ\mathcal{U}_1(\rho_0) +p(1-p)\rho_2 + p\rho_1\\
    &= \cdots \\ 
    &= (1-p)^L\mathcal{U}_{\textup{UCC}}(\rho_0) + \sum_{i=1}^L p(1-p)^{i-1} \rho_i, 
    \end{aligned}
\end{equation*}
where, for each $1 \leq i \leq L$, $\rho_i$ is a quantum state defined as 
\begin{equation*}
    \rho_i \coloneqq (\mathcal{E}_L\circ\mathcal{U}_L) \circ (\mathcal{E}_{L-1}\circ\mathcal{U}_{L-1}) \circ \cdots (\mathcal{E}_{i+1}\circ\mathcal{U}_{i+1}) \circ \Delta_i \circ \mathcal{U}_i \circ \mathcal{U}_{i-1} \circ \cdots \circ \mathcal{U}_2 \circ \mathcal{U}_1(\rho_0)
\end{equation*}
with the completely depolarizing channel $\Delta_i$ on the system on which $\mathcal{E}_i$ acts. 
Then, we have the following chain of inequalities:
\begin{equation}~\label{eq:eps_tot}
\begin{aligned}
    \epsilon_{\textup{tot}}
    &= \|\mathcal{N}_{\textup{UCC}}(\rho_0) - \mathcal{U}_{\textup{UCC}}(\rho_0)\|_1 \\
    &= \left\|\left[(1-p)^L - 1\right]\mathcal{U}_{\textup{UCC}}(\rho_0) + \sum_{i=1}^L p(1-p)^{i-1} \rho_i \right\|_1 \\ 
    &\leq 1 - (1-p)^L +  \sum_{i=1}^L p(1-p)^{i-1} \\ 
    &= 2\left(1 - (1-p)^L\right)\\ 
    &\sim L\times O(\epsilon^2).
\end{aligned}
\end{equation}
Therefore, since $p = O(\epsilon^2)$ for the averaged Clifford+$T$ approximation error where $\epsilon$ is Clifford+$T$ approximation error for a single-qubit $z$-rotation [Eq.~\eqref{eq: def of single SK error}], the total error $\epsilon_{\textup{tot}}$ caused by Clifford+$T$ approximation error of each gate can be modeled by a function 
\begin{equation*}
    2\left(1 - \left(1-O(\epsilon^2)\right)^L\right).
\end{equation*}

Since the total Clifford+$T$ approximation error $\epsilon_{\textup{tot}}$ above is given in the trace norm, it is also easy to estimate the error of an expectation value for some observable.
Suppose that we evaluate the energy of a given Hamiltonian $H$ with the UCC ansatz. 
When the ansatz circuit is ``noisy'', or approximated, due to the Clifford+$T$ decomposition, the energy $\tilde{E}$ we obtain under such situation is given by
\begin{equation*}
E_{\text{approx}} \coloneqq \tr(\mathcal{N}_{\textup{UCC}}(\rho_0)H). 
\end{equation*}
The difference of the expectation values of $H$ between the ideal ansatz state,
\begin{equation*}
    E_{\text{ideal}} = \tr(\mathcal{U}_{\textup{UCC}}(\rho_0)H), 
\end{equation*}
and the approximated (Clifford+$T$ decomposed) state is bounded by 
\begin{equation}~\label{eq:SK_energy}
    \begin{aligned}
    |E_{\text{approx}} - E_{\text{ideal}}| 
    &= \left|\tr(\mathcal{N}_{\textup{UCC}}(\rho_0)H) - \tr(\mathcal{U}_{\textup{UCC}}(\rho_0)H)\right| \\
    &\leq \|\mathcal{N}_{\textup{UCC}}(\rho_0) - \mathcal{U}_{\textup{UCC}}(\rho_0)\|_1\|H\| \\
    &\leq 2\|H\|\left(1 - \left(1-O(\epsilon^2)\right)^L\right).
    \end{aligned}
\end{equation}
The H{\"o}lder's inequality~\cite{watrous2018} is used at the first inequality. 
The second inequality follows from Eq.~\eqref{eq:eps_tot}. 
Hence, the error of the energy obtained by the Clifford+$T$ decomposition is expected to depend also on $\epsilon^2$ as shown in Eq.~\eqref{eq:SK_energy}. 

Note that the analysis conducted in this subsection is applicable not only to the UCC ansatz, but to a broader class of parametrized quantum circuits, in which only $z$-rotation gates are parametrized. 
Since Clifford+$z$-rotation gates form a universal gate set, we can convert any given parametrized circuit into this form. 
Thus, in theory, we may use this analysis for an arbitrary ansatz circuit. 
Practically, this method can be helpful when the Clifford+$z$-rotation expression of a given ansatz circuit can be easily obtained. 

\begin{remark}[Naive upper bound by triangle inequality]
\label{remark:triangle}
    If we do not use the averaging method, we only have a looser upper bound using the triangle inequality. 
    Here, we show the derivation of the naive bound. 

    Let us consider a UCC ansatz circuit $U_{\mathrm{UCC}} = U_LU_{L-1} \cdots U_{1}$, where each $U_i$ consists of Clifford gates and only one $z$-rotation gate. 
    Without the averaging method, the resulting circuit by Clifford + $T$ approximation can be written as 
    \begin{equation*}
        \tilde{U}_{\mathrm{UCC}} = \tilde{U}_L \circ \tilde{U}_{L-1} \circ \cdots \circ \tilde{U}_{1}, 
    \end{equation*}
    where for each $i = 1,2,\ldots, L$, $\tilde{U}_{i}$ only consists of Clifford gates and satisfies 
    \begin{equation*}
        \|\tilde{U}_{i} - U_{i}\| \leq \epsilon. 
    \end{equation*}
    In this case, by using the triangle inequality, the total approximation error $\tilde{\epsilon}_{\mathrm{tot}}$ is evaluated as 
    \begin{align*}
            \tilde{\epsilon}_{\mathrm{tot}} 
            &= \left\|\tilde{U}_{\mathrm{UCC}} \rho_0\tilde{U}_{\mathrm{UCC}}^\dagger - U_{\mathrm{UCC}} \rho_0 U_{\mathrm{UCC}}^\dagger \right\|_1 \\ 
            &= \left\|\tilde{U}_L\tilde{U}_{L-1} \cdots \tilde{U}_{1} \rho_0 \tilde{U}_1^\dagger \cdots \tilde{U}_{L-1}^\dagger \tilde{U}_{L}^\dagger - U_LU_{L-1} \cdots U_{1} \rho_0 U_1^\dagger \cdots U_{L-1}^\dagger U_{L}^\dagger \right\|_1 \\ 
            &=  \Bigg\|\tilde{U}_L\tilde{U}_{L-1} \cdots \tilde{U}_{1} \rho_0 \tilde{U}_1^\dagger \cdots \tilde{U}_{L-1}^\dagger \tilde{U}_{L}^\dagger 
            -U_L\tilde{U}_{L-1} \cdots \tilde{U}_{1} \rho_0 \tilde{U}_1^\dagger \cdots \tilde{U}_{L-1}^\dagger U_{L}^\dagger \\ 
            &\quad +U_L\tilde{U}_{L-1} \cdots \tilde{U}_{1} \rho_0 \tilde{U}_1^\dagger \cdots \tilde{U}_{L-1}^\dagger U_{L}^\dagger 
            -U_LU_{L-1}\tilde{U}_{L-2} \cdots \tilde{U}_{1} \rho_0 \tilde{U}_1^\dagger \tilde{U}_{L-2}\cdots U_{L-1}^\dagger U_{L}^\dagger \\
            &\quad +\cdots \\
            &\quad +U_LU_{L-1} \cdots \tilde{U}_{1} \rho_0 \tilde{U}_1^\dagger \cdots U_{L-1}^\dagger U_{L}^\dagger 
            - U_LU_{L-1} \cdots U_{1} \rho_0 U_1^\dagger \cdots U_{L-1}^\dagger U_{L}^\dagger \Bigg\|_1 \\ 
            &\leq 
            \left\|\tilde{U}_L \tilde{\rho}_{L-1}\tilde{U}_L^\dagger - U_L \tilde{\rho}_{L-1} U_L^\dagger \right\| \\ 
            &\quad + \sum_{i=0}^{L-2} \left\|(U_L\cdots U_{i+2})\left( \tilde{U}_{i+1} \tilde{\rho}_{i} \tilde{U}_{i+1}^\dagger - U_{i+1} \tilde{\rho}_{i} U_{i+1}^\dagger\right)(U_L\cdots U_{i+2})^\dagger \right\|_1, 
        \end{align*}
        where
    \begin{equation*}
        \tilde{\rho}_i \coloneqq 
        \begin{cases}
            \rho_0 & i = 0 \\ 
             \tilde{U}_{i}\cdots \tilde{U}_1 \rho_0 \tilde{U}_{1}^\dagger \cdots \tilde{U}_{i}^\dagger & i = 1,2,\ldots,L-1. 
        \end{cases}
    \end{equation*} 
    Due to the unitary-invariance of the trace norm~\cite{watrous2018}, the last line can be rewritten as 
    \begin{equation}
        \label{eq:error_upper_bound_rewrite}
        \sum_{i=0}^{L-1} \left\| \tilde{U}_{i+1} \tilde{\rho}_{i} \tilde{U}_{i+1}^\dagger - U_{i+1} \tilde{\rho}_{i} U_{i+1}^\dagger \right\|_1. 
    \end{equation}
    Each term in Eq.~\eqref{eq:error_upper_bound_rewrite} can be further analyzed as 
    \begin{align*}
        &\left\|\tilde{U}_{i+1} \tilde{\rho}_{i} \tilde{U}_{i+1}^\dagger - U_{i+1} \tilde{\rho}_{i} U_{i+1}^\dagger \right\|_1 \\  
        &= \left\|\tilde{U}_{i+1} \tilde{\rho}_{i} \tilde{U}_{i+1}^\dagger - U_{i+1} \tilde{\rho}_{i} \tilde{U}_{i+1}^\dagger + U_{i+1} \tilde{\rho}_{i} \tilde{U}_{i+1}^\dagger - U_{i+1} \tilde{\rho}_{i} U_{i+1}^\dagger \right\|_1 \\ 
        &\leq \left\|\tilde{U}_{i+1} \tilde{\rho}_{i} \tilde{U}_{i+1}^\dagger - U_{i+1} \tilde{\rho}_{i} \tilde{U}_{i+1}^\dagger  \right\|_1 +  \left\|U_{i+1} \tilde{\rho}_{i} \tilde{U}_{i+1}^\dagger - U_{i+1} \tilde{\rho}_{i} U_{i+1}^\dagger \right\|_1 \\ 
        & \leq \|\tilde{U}_{i+1} - U_{i+1} \| \|\tilde{\rho}_i\|_1 \|\tilde{U}_{i+1}^\dagger\| + \|U_{i+1}\| \|\tilde{\rho}_i\|_1 \|\tilde{U}_{i+1}^\dagger - U_{i+1}^\dagger \| \\ 
        &= \|\tilde{U}_{i+1} - U_{i+1} \| + \|\tilde{U}_{i+1}^\dagger - U_{i+1}^\dagger \| \\ 
        &= 2\|\tilde{U}_{i+1} - U_{i+1} \| \\ 
        &\leq 2\epsilon. 
    \end{align*}
    In the first inequality, we used the triangle inequality, and in the second inequality, we use the fact 
    \begin{equation*}
        \|ABC\|_1 \leq \|A\| \|B\|_1 \|C\|
    \end{equation*}
    for linear operators $A,B,C$~\cite{watrous2018}.
    In the second inequality, we used the fact that 
    \begin{equation*}
        \|U\| = 1
    \end{equation*}
    for any unitary operator $U$. 
    The last equality follows from 
    \begin{equation*}
        \|A\| = \|A^\dagger\| 
    \end{equation*}
    for linear operator $A$~\cite{watrous2018}. 
    In summary, with this approach, we have 
    \begin{equation}
        \label{eq:naive_epsilon}
        \tilde{\epsilon}_{\mathrm{tot}} \leq \sum_{i=1}^L 2\epsilon = 2L\epsilon. 
    \end{equation}
    While the averaging method lead to an upper bound with the scaling of $\sim L\times O(\epsilon^2)$, this naive method yields the linear upper bound $2L\epsilon$. 

    Now, we evaluate the expectation value of $H$ under this approximated ansatz state, 
    \begin{equation}
        \tilde{E}_{\text{approx}} = \tr\left(\tilde{U}_{\text{UCC}}\rho_0 \tilde{U}_{\text{UCC}}^\dagger H\right). 
    \end{equation}
    By using the bound~\eqref{eq:naive_epsilon}, in a similar matter to Eq.~\eqref{eq:SK_energy}, 
    we have  \begin{equation}~\label{eq:SK_energy_no_average}
    \begin{aligned}
    | \tilde{E}_{\text{approx}} - E_{\text{ideal}}| 
    &= \left|\tr\left(\tilde{U}_{\text{UCC}}\rho_0 \tilde{U}_{\text{UCC}}^\dagger H\right) - \tr\left(U_{\text{UCC}}\rho_0 U_{\text{UCC}}^\dagger H\right)\right| \\
    &\leq \left\|\tilde{U}_{\text{UCC}}\rho_0 \tilde{U}_{\text{UCC}}^\dagger - U_{\text{UCC}}\rho_0 U_{\text{UCC}}^\dagger\right\|_1\|H\| \\
    &\leq 2\|H\|L\epsilon.
    \end{aligned}
\end{equation}
\end{remark}

\subsection{Algorithm of $T$-Count Fixed Gate Decomposition}~\label{sec:algorithm}
In this section, we present our algorithm to obtain a decomposition of a given single-qubit $z$-rotation \textit{with a fixed number of $T$ gates}. 
In the successive section (Sec.~\ref{subsec:numerical_experiments}), we perform numerical simulations by using this algorithm to figure out the performance of the UCC ansatz in the Pre-FTQC era. 
We expect that our algorithm is useful when one wants to obtain a good Clifford+$T$ decomposition when the available number of $T$ gates is pre-determined, which should be highly plausible in Pre-FTQC. 
Our algorithm is based on the one proposed in Ref.~\cite{Ross2016}, which yields a decomposition of a given $z$-rotation with the \textit{minimal number of $T$ gates} within a \textit{fixed accuracy}. 

First, we introduce notations and background materials for the algorithm. 
Let $\omega$ denote a complex number $\mathrm{e}^{i\tfrac{\pi}{4}}$. 
We define the following rings, which are given as extensions of the ring of the integers $\mathbb{Z}$. 
\begin{definition}
The ring $\mathbb{Z}[\omega]$ is defined by 
\begin{equation*}
    \mathbb{Z}[\omega] \coloneqq \left\{a\omega^3 + b\omega^2 + c\omega + d \middle| a,b,c,d \in \mathbb{Z}\right\}. 
\end{equation*}
The ring $\mathbb{D}$ is defined by 
\begin{equation*}
    \mathbb{D} \coloneqq \left\{\dfrac{a}{2^k} \middle|a \in \mathbb{Z}, k=0,1,2,\ldots \right\}. 
\end{equation*}
The ring $\mathbb{D}[\omega]$ is defined by 
\begin{equation*}
    \mathbb{D}[\omega] \coloneqq \left\{a\omega^3 + b\omega^2 + c\omega + d \middle| a,b,c,d \in \mathbb{D}\right\}. 
\end{equation*}
\end{definition} 

\begin{definition}
Let $t \in \mathbb{D}[\omega]$ and $k\in \mathbb{N}.$ 
If $\sqrt{2^k}t \in \mathbb{Z}[\omega]$, then we say that $k$ is a \emph{denominator exponent} of $t$. 
The smallest such $k \geq 0$ is the \emph{least denominator exponent} of $t$. 
\end{definition}
Now, let us describe our gate decomposition algorithm. 
Due to Lemmas 7.2 and 7.3 of Ref.~\cite{Ross2016}, the task of finding a good approximation of a given $z$-rotation $R_z(\theta)=e^{-i\frac{\theta}{2}Z}$ within a given number $N_T$ of $T$-gates can be reduced to the following searching problem. 
\begin{problem}
Let $\theta \in [0,2\pi)$ be a given angle of $z$-rotation, and let $N_T$ be a positive integer representing the maximum number of $T$ gates available. 
Find a pair of $t,u \in \mathbb{D}[\omega]$ maximizing $\mathrm{Re}(u\mathrm{e}^{-i\theta/2})$
under the following constraints: 
\begin{enumerate}[(a)]
    \item $tt^\dagger + uu^\dagger = 1$,  
    \item and $u$ has a denominator exponent $\left\lfloor\tfrac{N_T}{2}\right\rfloor + 1$. 
\end{enumerate}
\end{problem}
Once a pair of $(t,u)$ are obtained, the unitary matrix 
\begin{equation*}
U(t,u) \coloneqq \left(
\begin{array}{cc}
    u & -t^\dagger \\
    t & u^\dagger
\end{array}
\right)
\end{equation*}
serves as a good approximation of the given $z$-rotation $R_z(\theta)$. 
There exists an efficient algorithm to obtain the exact decomposition of $U(t,u)$ into $H$, $S$, and $T$ gates~\cite{Kliuchnikov2013}. 
In this exact decomposition algorithm, the number of $T$ gates used is given by $2k-2$ with the least denominator exponent $k$ of $u$ when $k>0$, and $T$-count is zero when $k=0$. 
Simple calculation shows that Clifford+$T$ approximation error is indeed given by  
\begin{equation}
\label{eq: def of single SK error}
     \epsilon \coloneqq \|U(t,u) - R_z(\theta)\| = \sqrt{2(1 -\mathrm{Re}(u\mathrm{e}^{-i\theta/2}))}, 
\end{equation}
where $\|\cdot\|$ is the operator norm. 

Now, we show an algorithm to have the best approximate decomposition of a given $z$ rotation within a given cost of $T$-gates.

\begin{algorithm}~\label{alg:ross}
Suppose that an angle $\theta$, the maximum number $N_{T}$ of $T$ gates, and an estimate of Clifford+$T$ approximation error for each gate $\tilde{\epsilon} > 0$ are given.   
By using Lemma 7.4 and Proposition 5.21 of Ref.~\cite{Ross2016}, we can efficiently enumerate the candidates of $u \in \mathbb{D}[\omega]$ with the denominator exponent $\left\lfloor\tfrac{N_T}{2}\right\rfloor + 1$ that satisfy $tt^\dagger + uu^\dagger = 1$ and  $\mathrm{Re}(u\mathrm{e}^{-i\theta/2}) \geq 1 - \tilde{\epsilon}^2/2$. 
Let $\mathbb{U}_{\textup{cand}}$ be the set of such candidates $u$. 
If no such $u$ was found, that is, $\mathbb{U}_{\textup{cand}} = \emptyset$, we set $\tilde{\epsilon} \leftarrow 2\tilde{\epsilon}$, and try with this new error estimate. 
 
\begin{enumerate}[(i)]
\item Set $\epsilon_{\textup{tmp}} = 2$, and for each candidate $u \in \mathbb{U}_{\textup{cand}}$, we run the following procedure: \\
Use Theorem 6.2 of Ref.~\cite{Ross2016} to solve $tt^\dagger = 1 - uu^\dagger$ with respect to $t$. 
If there is no solution or $\epsilon = \sqrt{2 - 2\mathrm{Re}(u\mathrm{e}^{-i\theta/2})} \geq \epsilon_{\textup{tmp}}$, then skip to the next candidate in $\mathbb{U}_{\textup{cand}}$. 
Otherwise, set 
\begin{align}
    t_{\textup{tmp}} \leftarrow t, ~~ 
    u_{\textup{tmp}} \leftarrow u, ~~
    \epsilon_{\textup{tmp}} \leftarrow \epsilon
\end{align}
and continue with the next $u$. 
\item After examining all the candidates, 
use the exact synthesis algorithm of Ref.~\cite{Kliuchnikov2013} to find decomposition of $U(t_{\textup{tmp}},u_{\textup{tmp}})$ into $H$, $S$, and $T$ gates. 

\item Output the circuit $U(t_{\textup{tmp}},u_{\textup{tmp}})$ and Clifford+$T$ approximation error $\epsilon_{\textup{tmp}}$, and stop. 
\end{enumerate}
\end{algorithm}

\begin{remark}[Error Estimate $\tilde{\epsilon}$]
Algorithm~\ref{alg:ross} searches all the Clifford+$T$-gate decompositions within estimate error $\tilde{\epsilon}$, and then outputs the one with the smallest error. 
Thus, in theory, a sufficiently large $\tilde{\epsilon}$, say, $\tilde{\epsilon} = 2$, guarantees the smallest-error decomposition is obtained through this algorithm. 
Practically, smaller $\tilde{\epsilon}$ may be preferable in terms of (classical) computational efficiency since the search region $\mathbb{U}_{\textup{cand}}$ is smaller when we take smaller $\epsilon$. 
Hence, we consider inputting some small enough error estimate or desired accuracy as an initial $\tilde{\epsilon}$, and updating $\tilde{\epsilon} \leftarrow 2\tilde{\epsilon}$ when no decomposition is obtained. 
Since the algorithm returns the smallest-error decomposition anyway, the choice of the initial estimated error will not affect the validity of the algorithm. 
\end{remark}

\begin{remark}[Comparison with the algorithm in Ref.~\cite{Ross2016}]
The algorithm proposed in Ref.~\cite{Ross2016} provides a Clifford+$T$-gate decomposition with the minimum $T$-count under given Clifford+$T$ approximation error $\epsilon$. 
Thus, in that algorithm, one enumerates the candidate of $u$ with the smallest possible denominator exponent $k$, say $k=0$, that satisfies $tt^\dagger + uu^\dagger = 1$ and $\mathrm{Re}(u\mathrm{e}^{-i\theta/2}) \geq 1 - \epsilon^2$ with the fixed given $\epsilon$. 
If no such pair $(t,u)$ was found, then one increments the denominator exponent $k$ by $1$, and repeat the same procedure until such a pair is found. 
Once a pair of $(t,u)$ is found, then one produces the corresponding Clifford+$T$-gate decomposition using the exact synthesis algorithm~\cite{Kliuchnikov2013}.
The primary difference between these two algorithms is that we fix $N_T$ and seek the best $\epsilon$ while they fixed $\epsilon$ and sought the best $N_T$. 
\end{remark}

\begin{remark}[Necessity of Prime Factorization]
Similar to the algorithm by Ross and Selinger~\cite{Ross2016}, our algorithm demands prime factorization of some integer when we solve $tt^\dagger + uu^\dagger = 1$. 
Although factoring is in \NP, for which a computationally efficient algorithm has not been found, factoring in this algorithm can be done fast practically by the brute-force method since the integer which we want to perform factorization is relatively small, as also indicated in Ref.~\cite{Ross2016}. 
\end{remark}

\subsection{Numerical experiments for the UCC ansatz}~\label{subsec:numerical_experiments}
Here we numerically illustrate the theoretical analysis on the average Clifford+$T$ approximation error above by actually decomposing the quantum circuit of the UCC ansatz using the algorithm described in Sec.~\ref{sec:algorithm}.

Our setup for numerical simulations is as follows.
We consider a wide variety of molecules shown in Table~\ref{tab: geometries}.
We adopt the STO-3G minimal basis set, and construct the fermionic second-quantized Hamiltonians for electrons with the Hartree-Fock orbitals by using the numerical libraries PySCF~\cite{Sun2018} and OpenFermion~\cite{Mcclean2017}.
We then employ Jordan-Wigner transformation~\cite{Jordan1928} to map the fermionic Hamiltonians into qubit Hamiltonians~\cite{Mcardle2018, Cao2018}.
We consider decomposing the unitary coupled cluster singles and doubles (UCCSD) ansatz circuit for those molecules, which is explicitly defined in Appendix~\ref{sec:AppendixA}, into Clifford+$T$-gate circuit by using the algorithm in Sec.~\ref{sec:algorithm}. 
In the simulation, we do not optimize the parameters in the UCCSD ansatz.
We instead use the amplitudes of the coupled cluster singles and doubles (CCSD)~\cite{Bartlett2007} computed on classical computers with PySCF~\cite{Sun2018} as parameters of the UCCSD ansatz.
Hereafter, we call this state the \textit{UCCSD' state}. 
We generate Clifford+$T$-gate decomposition of the UCC' ansatz for each molecule with various $T$-count $N_T$, which stands for the maximum number of $T$-gates used in Algorithm 1 in Sec.~\ref{sec:algorithm}.
We consider $22 \leq N_T \leq 50$ in the simulation.
All simulations for quantum circuits are performed by the high-speed quantum circuit simulator Qulacs~\cite{Suzuki2020}.
We do not include any noise for quantum circuit simulations. 

\begin{table}[] 
 \caption{Geometries of molecules used in the numerical simulations. ``$(\mathrm{X}, (x,y,z))$'' denotes three dimensional coordinates $x,y,z$ of atom X in units of \AA. 
 \label{tab: geometries}
 }
 \begin{tabular}{c|c|l}
 \hline \hline
 \textbf{Molecule} & \textbf{\# of Qubits} &\textbf{Geometry}  \\ \hline
 \ce{H2} &4 & (H, (0, 0, 0)), (H, (0, 0, 0.735)) \\
 \ce{H}$_i$  & $2i$ & (H,($0,0,j-1$)) for the $j$th \ce{H} atom ($j = 1,2,\ldots,i$). \\($i = 4,6,8,10$) & & \\
 \ce{H2O} &14 & (O, (0, 0, 0.137)), (H, (0,0.769,-0.546)), (H, (0,-0.769,-0.546)) \\
 \ce{LiH} &12 & (Li, (0, 0, 0)), (H, (0, 0, 1.548)) \\
 \ce{BeH2} &14 & (Be, (0, 0, 0)), (H, (0, 0, -1.304)), (H, (0, 0, 1.304)) \\
 \ce{H2S} &22 & (S, (0, 0, 0.1030)), (H, (0, 0.9616, -0.8239)), \\ 
 && (H, (0, -0.9616, -0.8239)) \\
\ce{CH4} &18 & (C, (0, 0, 0)), \\ && (H, (0.6276, 0.6276, 0.6276)), (H, (0.6276, -0.6276, -0.6276)), \\
 & & (H, (-0.6276, 0.6276, -0.6276)), (H, (-0.6276, -0.6276, 0.6276)) \\
 \ce{C2H2} &24 & (C, (0, 0, 0.6013)), (C, (0, 0, -0.6013)), \\&&(H, (0, 0, 1.6644)), (H, (0, 0, -1.6644))
\\ 
\ce{CO} &20 & (C, (0, 0, 0)), (O, (0, 0, 1.128)) \\
\ce{N2} &20 & (N, (0, 0, 0)), (N, (0, 0, 1.5)) \\
\ce{NH3} &16 & (N, (0,0,0.149)), (H, (0,0.947,-0.348)),\\ && (H, (0.821,-0.474,-0.348)), (H, (-0.821,-0.474,-0.348)) \\
\hline \hline
 \end{tabular}
\end{table}

First, we examine the Clifford+$T$ approximation error analysis proposed in Sec.~\ref{sec:average_SK}. 
We calculate the energy expectation values of the approximate UCCSD' state under Clifford+$T$ decomposition with the single-gate $T$-count $N_T$. 
Recalling that the average effects of Clifford+$T$ approximation error on the UCC ansatz is given by Eq.~\eqref{eq:SK_energy}, 
we model the energy expectation value for a given $N_T$ as 
\begin{equation}~\label{eq:model}
    E_{\text{ideal}} + 2E_{\text{ideal}} \left(1 - \left( 1 - c \cdot  (\epsilon(N_T))^2 \right)^L \right), 
\end{equation}
where 
$\epsilon(N_T) \coloneqq 10^{-\tfrac{N_T}{10}}$ is the single-gate Clifford+$T$ approximation error, 
$L$ is the total number of $z$ rotations in the UCCSD ansatz, 
$E_{*}$ is the energy expectation value for the original UCCSD ansatz (not decomposed into Clliford+$T$), and $c$ is a fitting parameter. 
The functional form of the single-gate Clifford+$T$ approximation error $10^{-\tfrac{N_T}{10}}$ is assumed by following the numerical observation in Ref.~\cite{Ross2016}.
For each molecule, the data of the energy expectation values versus $N_T$ is fitted by performing the (non-linear) regression to this model~\eqref{eq:model}. 
We take the range of  $32 \leq N_T\leq 50$ for the fitting to see the accuracy of the model for relatively large $N_T$. 
The results for $\ce{C_2H_2}$ and $\ce{NH_3}$ molecules, which exhibit representative behaviors, are shown in Fig.~\ref{fig:energy}.
The whole results for all molecules in Table~\ref{tab: geometries}  are given in Appendix~\ref{sec:AppendixB}. 
As one may see, the model~\eqref{eq:model} explains the total Clifford+$T$ approximation error well while the previous error estimate by the triangle inequality gives much worse upper bound of the energy (see the captions).
This is an illustration of the validity of the average error analysis presented in Sec.~\ref{sec:average_SK}.
We also compare our results with the naive estimation using the triangle inequality~\eqref{eq:SK_energy_no_average} (see the caption and subcaptions of Fig.~\ref{fig:energy}). 
Note that while $\|H\|$ cannot be obtained when a given quantum system is large, we here use the true value of the ground energy. 
The comparison indeed shows that our model gives a more precise estimation of energy expectation values. 

\begin{figure}[htbp]
  \begin{subfigure}[b]{0.48\linewidth}
    \includegraphics[keepaspectratio, width=\linewidth]{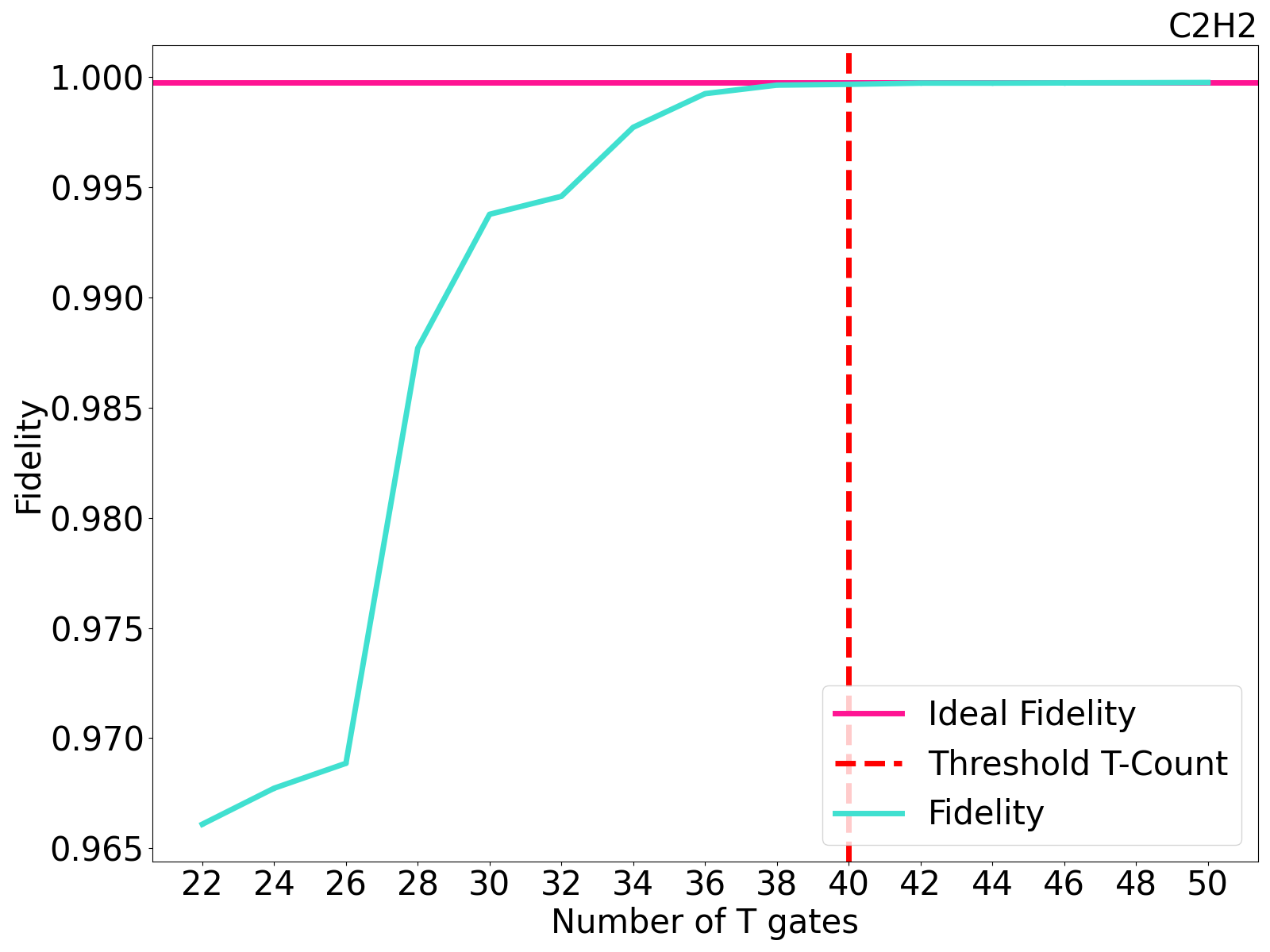}
    \subcaption{Fidelity of the UCCSD' states of \ce{C_2H_2} to the exact ground state.}
  \end{subfigure}
  \begin{subfigure}[b]{0.48\linewidth}
    \includegraphics[keepaspectratio, width=\linewidth]{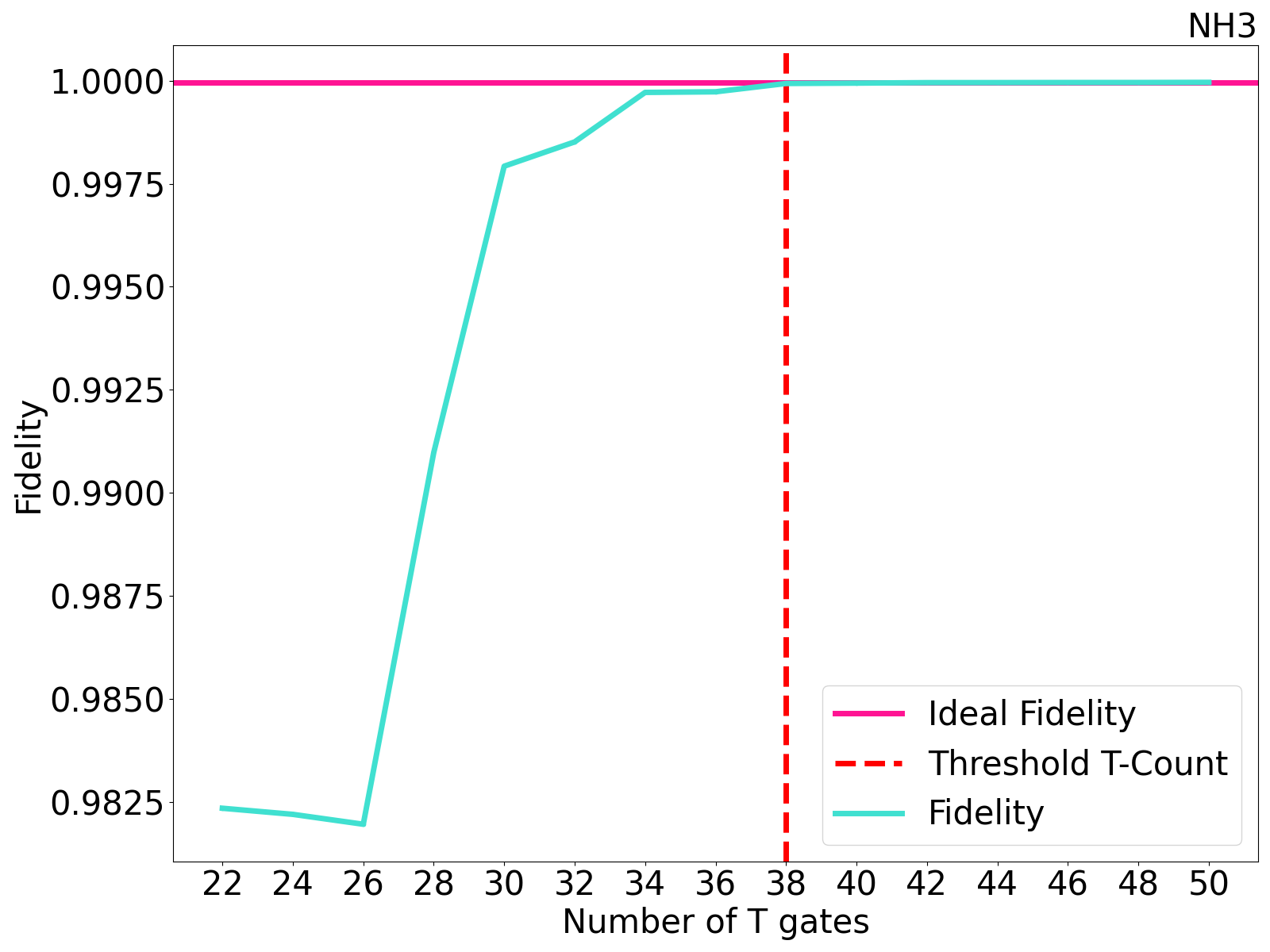}
    \subcaption{Fidelity of the UCCSD' states of \ce{NH3} to the exact ground state.}
  \end{subfigure}
  \caption{The fidelity between the exact ground state of the Hamiltonian and the Clifford+$T$-gate decomposed UCCSD' states with $T$-gate budget $22 \leq N_T \leq 50$. In each panel, the blue curve represents the fidelity between the exact ground state and the UCCSD' state that is decomposed by Algorithm~\ref{alg:ross}. The pink horizontal line indicates the fidelity between the exact ground state and the true UCCSD' state obtained by the UCCSD ansatz without Clifford+$T$ decomposition. The orange vertical line represents the $T$-count threshold that achieves Eq.~\eqref{eq:rel_error}.}
    \label{fig:fidelity}
\end{figure}

\begin{figure}[htbp]
    \centering
    \includegraphics[keepaspectratio, width=0.8\linewidth]{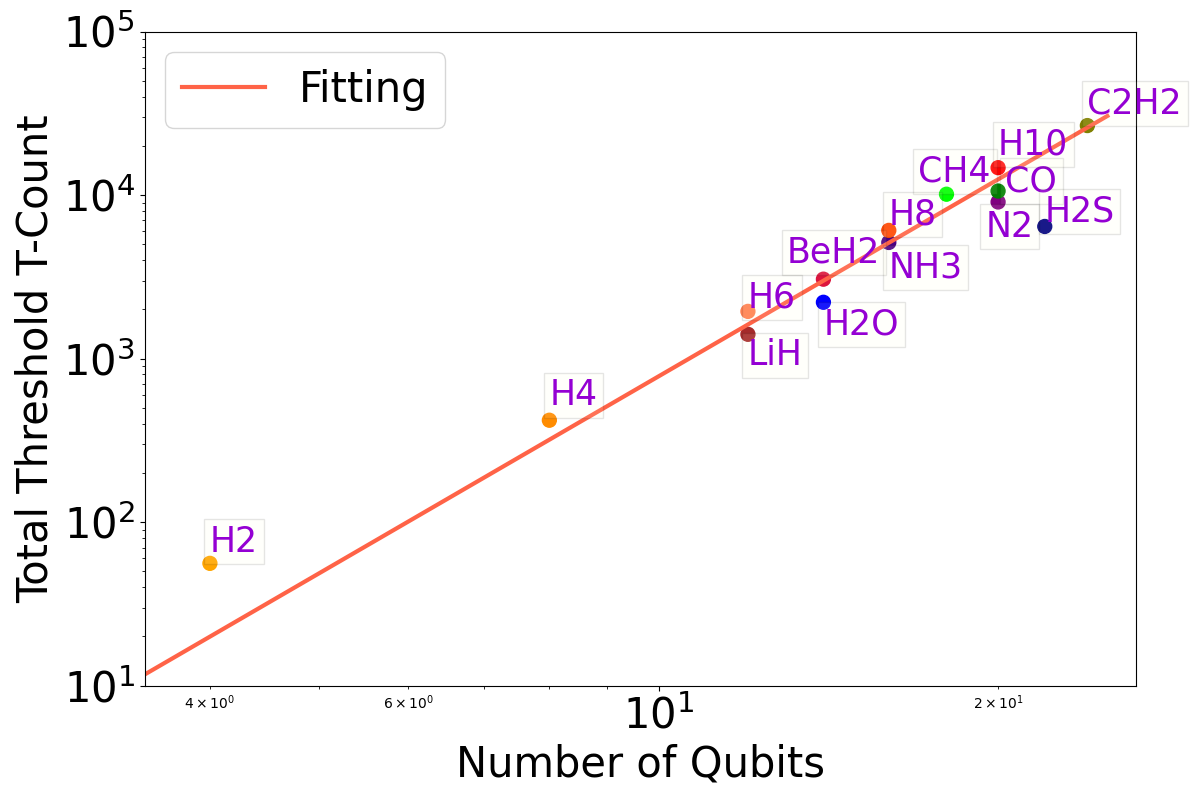}
  \caption{Number of qubits vs \textit{Total} threshold $T$-count. 
  The red curve is the fitting of the data of the total threshold $T$-count versus the number of qubits by a quartic function $an^4$ with the fitting result $a = 0.07795(1)$, where $n$ is the number of qubits. 
  }
  \label{fig:fitting_threshold}
\end{figure}

Next, we evaluate the threshold for the $T$-count $N_T$ with which we can guarantee sufficient accuracy to represent the UCCSD' state.
In Fig.~\ref{fig:fidelity}, we plot the fidelity between the exact ground state of the Hamiltonian and the approximate (Clifford+$T$-gate-decomposed) UCCSD' state with respect to the $T$-count. 
As expected, the larger $N_T$ results in higher fidelity.
We determine the threshold for $T$-count to represent the UCCSD' state with high accuracy by requiring that the following quantity,  
\begin{equation}~\label{eq:rel_error}
    \delta_{\textup{rel}} \coloneqq \dfrac{|F(N_T) - F_{\textup{ideal}}|}{F_{\textup{ideal}}},
\end{equation}
satisfies $\delta_{\textup{rel}} < 10^{-4}$,
where $F(N_T)$ is the fidelity between the approximated UCCSD' state and the exact ground state of the Hamiltonian and $F_{\textup{ideal}}$ is the fidelity between the original (not-decomposed)  UCCSD' state and the exact ground state.
$\delta_{\textup{del}}$ means the relative error of the fidelity with respect to the exact ground state of the Hamiltonian. 
The results for the \textit{total} threshold $T$-counts, ((the per-gate threshold $T$-counts obtained in Fig.~\ref{fig:fidelity}) $\times$ (the number of $z$-rotation gates)),
for various molecules are summarized in Fig.~\ref{fig:fitting_threshold}. 
Indeed, the total threshold $T$-count mostly increases quartically with respect to the number of qubits.
In Fig.~\ref{fig:fitting_threshold}, we also show the results of fitting of the total threshold $T$-count versus the number of qubits obtained by performing the linear regression using a quartic function $an^4$ with fitting parameter $a$ and the number $n$ of qubits. 
The fitting explains the data pretty well with fitting result $a = 0.07795(1)$. 
This $n^4$ dependence is natural when one considers there are $O(n^4)$ multi-qubit Pauli rotation gates in the UCCSD ansatz.

This result implies that the usage of variational quantum algorithms with a deep ansatz such as the UCC ansatz can be beneficial in the Pre-FTQC era. 
For example, consider the FeMo cofactor of the nitrogenase enzyme \ce{Fe7MoS9C}, known as \ce{FeMoco}.
There are two active space models considered conventionally for this molecule, one proposed by Reiher \textit{et al.}~\cite{Reiher2017} with the number of qubit $n = 104$ (RWSWT orbital) and the other one proposed by Li \textit{et al.}~\cite{Li2019} with $n = 154$ (LLDUC orbital). 
The current best resource estimation for the quantum phase estimation of \ce{FeMoco}, as far as we are aware of, is obtained using tensor hypercontraction by Lee \textit{et al.}~\cite{Lee2021}, and the estimate \textit{Toffoli} count is $5.3 \times 10^9$ for RWSWT \ce{FeMoco} and 
$3.2 \times 10^{10}$ for LLDUC \ce{FeMoco}. 
The most straightforward way to estimate $T$-count from Toffoli count is to simply transform the Toffoli gate into a quantum circuit using Clifford+$T$-gates, and this can be done using $7$ $T$ gates~\cite{nielsen_chuang_2010}. 
Or, recently, a magic-state factory construction based on the surface code was proposed~\cite{Gidney2019efficientmagicstate}, implying one Toffoli gate is roughly twice as costly as one $T$ gate. 
With these estimations, for both active space models, the $T$-count needed to perform the quantum phase estimation to \ce{FeMoco} is roughly $O(10^{10})\sim O(10^{11})$ at best. 
On the other hand, with our estimation, the $T$-count for realizing the UCCSD' state with the accuracy $\delta_{\textup{rel}} < 10^{-4}$ is $an^4 \approx 0.078\times (154)^4 \approx 4\times 10^7$ even for the larger active space of LLDUC \ce{FeMoco}, which is smaller than the estimate $T$-count for the quantum phase estimation by three orders of magnitude. 
In addition, the tensor hypercontraction demands enormous auxiliary qubits~\cite{Lee2021}; if we do not use auxiliary qubits, adopting the Trotter-based methods, we have the estimate $T$-count of roughly $O(10^{14})\sim O(10^{16})$. 
In this case, the $T$-count for the UCCSD' state is even smaller by seven-to-nine orders of magnitude. 
During the Pre-FTQC era, it is expected that we will experience time where we have $O(10^7)$ $T$ gates but still cannot perform the phase estimation algorithm due to the lack of a sufficient number of clean qubits, an possibly improved but still high error rate, and the limited number of $T$ gates. 
Our numerical results imply the Clifford+$T$ approximated UCCSD ansatz may be used to generate a good approximation of a desired state in such an era. 
Moreover, even in the \textit{post} Pre-FTQC era, where we are stepping into FTQC and the phase estimation algorithm is finally available, the Clifford+$T$ approximated UCCSD ansatz may be useful to generate a good initial state for the phase estimation considering the $T$-counts. 
Thus, the averaging method we proposed and our numerical analysis strongly suggest the utility of the combination of the Clifford+$T$ approximation and parametrized ansatz states in the Pre-FTQC era and beyond.

\begin{figure}[htbp]
  \begin{subfigure}[b]{0.48\linewidth}
    \includegraphics[keepaspectratio, width=\linewidth]{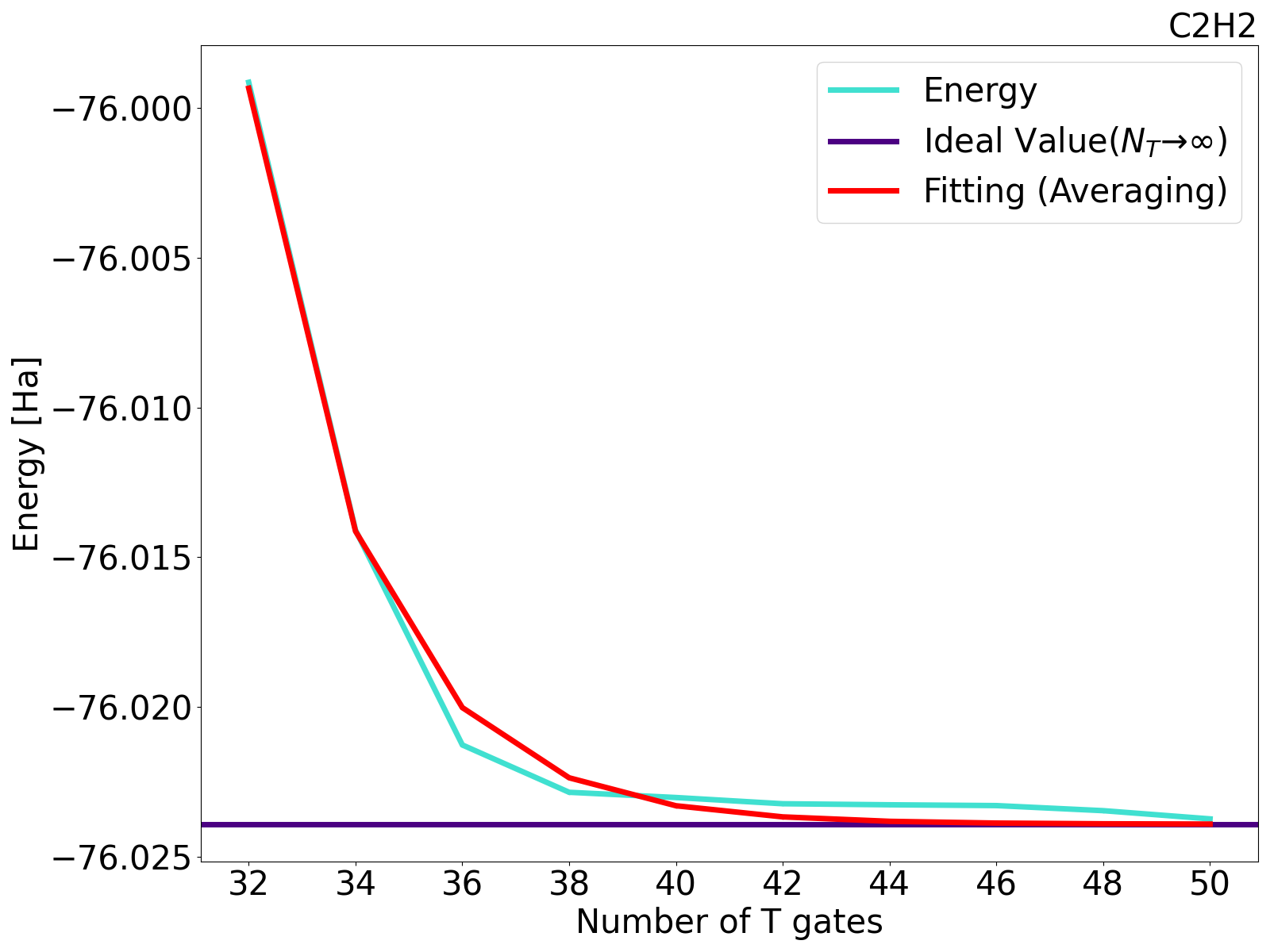}
    \subcaption{The energy expectation values of \ce{C_2H_2}. The fitting parameter is $c = 0.6106(2)$. On the other hand, the triangle inequality gives energy upper bound $-75.52\,\mathrm{Ha}$ at $N_T = 50$.}
  \end{subfigure}
  \begin{subfigure}[b]{0.48\linewidth}
    \includegraphics[keepaspectratio, width=\linewidth]{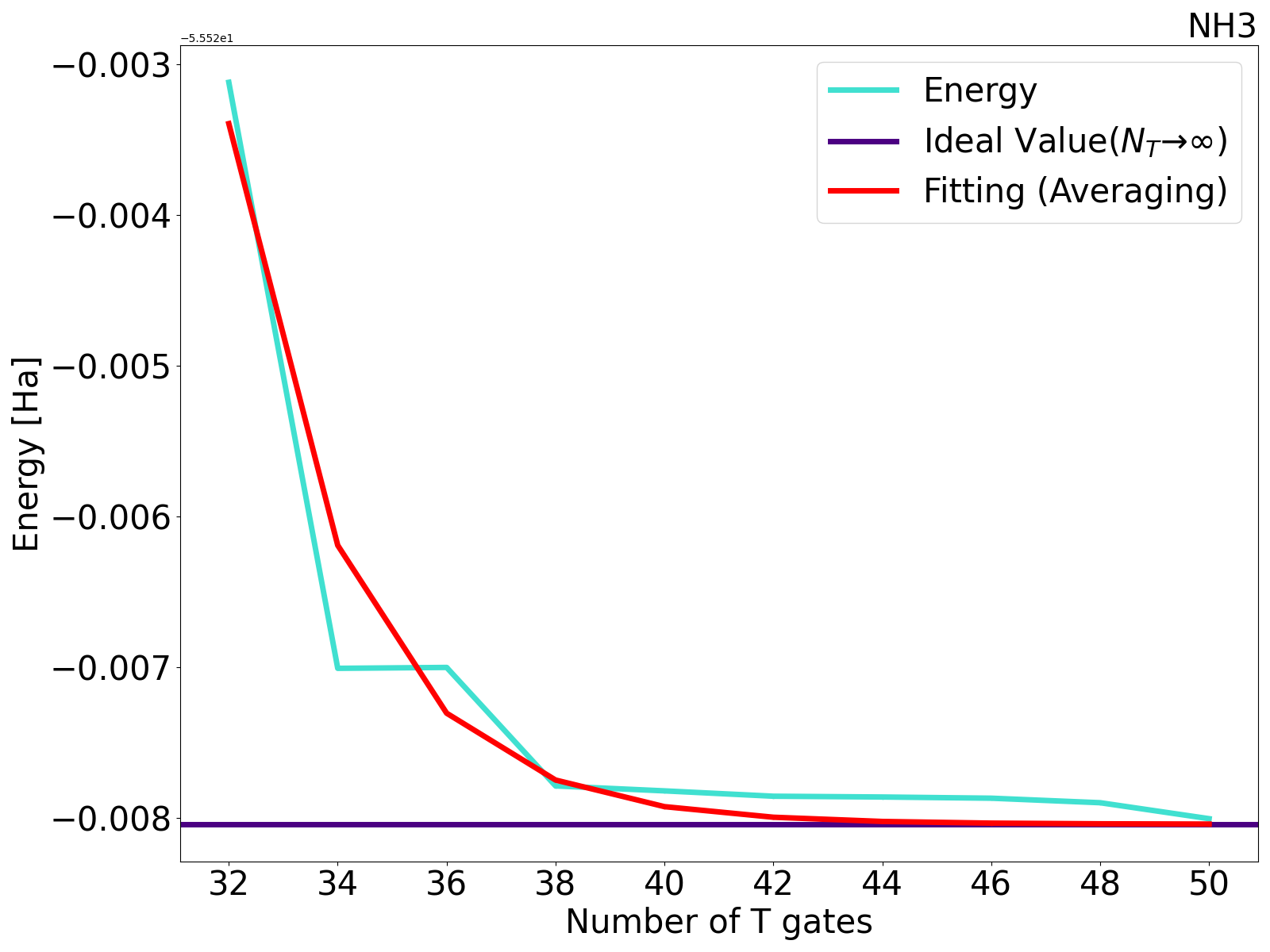}
    \subcaption{The energy expectation values of \ce{NH_3}. The fitting parameter is $c = 0.779(2)$. The triangle inequality gives energy upper bound $-55.45\,\mathrm{Ha}$ at $N_T = 50$.}
  \end{subfigure}
  \caption{Energy expectation values of Clifford+$T$-gate decomposed UCCSD' states with $T$-gate budget $32 \leq N_T \leq 50$. In each panel, the orange curve represents the expectation values of the UCCSD' state that are decomposed by Algorithm~\ref{alg:ross}. The purple horizontal line indicates the true UCCSD' state energy obtained by the UCCSD ansatz without Clifford+$T$ decomposition. The blue curve is the fitting of the orange curve modeled by Eq.~\eqref{eq:model}. 
  As indicated in each subcaption, the naive analysis by the triangle inequality is substantially rougher than our method.
  }
    \label{fig:energy}
\end{figure}

\newpage 
\section{Summary discussion, and outlook}~\label{sec:discussion}
In this work, we proposed a method of averaging the effects of the unitary Clifford+$T$ approximation error into the depolarizing noise when a given circuit is sufficiently large, which makes the analysis of the total Clifford+$T$ approximation error easier.
We apply this method to Unitary Coupled Cluster (UCC) ansatz. 
We constructed a Clifford+$T$-gate decomposition algorithm to provide the most accurate approximation of a given single-qubit $z$-rotation gate when the available number of $T$ gates is restricted. 
With this algorithm, we numerically verified our proposal for a wide variety of molecules, and based on the numerical simulations, we reveal the threshold $T$-counts to achieve desired accuracy. 
Hence, we shed a new light for the Pre-FTQC era, where only a restricted number of $T$ gates are practically available.
We believe that our proposal and analysis lead to more accurate resource estimation in the framework of Pre-FTQC. 
As a future direction, it would be nice if our algorithm and averaging method are applicable to another setup, \textit{e.g.}, quantum phase estimation algorithm, and estimate Clifford+$T$ approximation error or threshold $T$-count in that case. 
Moreover, it would also be interesting to investigate the performance of a Clifford+$T$ decomposed ansatz in a variational quantum algorithm, where the number of avalable $T$ gates is limited. 
In our numerical experiment, we compared the energy expectation values of the UCCSD' states with the energies computed by the classical CCSD method to verify the performance of our theoretical analysis. 
However, one may want to try to obtain a good estimate of the ground energy of a given molecule by a variational algorithm, in particular, the variational quantum eigensolver (VQE) with a Clifford+$T$ approximated ansatz with a limited number of $T$ gates in the Pre-FTQC era. 
By applying our analysis to the ground state, we believe that the resulting energy estimate obtained by the VQE with a Clifford+$T$ approximated ansatz behaves similarly to what we obtained in our numerical simulation. 
It would be great if this observation is verified in future research. 

\section*{Note added}
After posting the previous version of this manuscript on arXiv, during the preparation of this version of the manuscript, we noticed the appearance of a few papers relevant to our results. 
In Ref.~\cite{akibue_2024}, Akibue \textit{et al.} proved that gate approximation error becomes in the order of $O(\epsilon^2)$ when one performs gate approximation probabilistically, that is, randomly selecting an approximated gate from an appropriate set of candidates with appropriate probability distribution. 
In Ref.~\cite{yoshioka2024errorcraftingprobabilisticquantum}, Yoshioka \textit{et al.} proved that one can craft the remnant error of the probabilistic approximation of the target unitary into a Pauli channel, which is useful to apply the error countermeasure such as quantum error mitigation, while keeping the error $O(\epsilon^2)$.
They also numerically observed that the approximation error might drop $O(\epsilon^3)$ when one allows CPTP maps with ancillary qubits.

In this paper, contrary to their results, we do not consider probabilistic approximation or any countermeasure against error. 
Rather, we assume that when a given parametrized circuit is sufficiently large, the effect of gate approximation error for each gate will be averaged over the whole circuit. 
Under this setup, we observe that the total error can be model as the depolarizing channel with error $O(\epsilon^2)$. 
It would be interesting to see if there is a deeper connection between our analysis and results obtained by Refs.~\cite{akibue_2024,yoshioka2024errorcraftingprobabilisticquantum}. 

\section*{Acknowledgement}
The authors are grateful to Yasunori Lee for valuable discussions and comments on the manuscript.
We also thank Keita Kanno for fruitful discussions on the early stage of this work.
We would like to express our deep appreciation to Hayata Morisaki for providing us with useful codes of the Clifford+$T$ decomposition algorithm of Ref.~\cite{Ross2016}. 
The numerical simulations of this work are largely inspired by his implementation. 

KK is supported by a Mike and Ophelia Lazaridis Fellowship, the Funai Foundation, and a Perimeter Residency Doctoral Award. 

\bibliographystyle{quantum}
\bibliography{Tgate_Fixed_SK}

\newpage
\appendix 

\section{``Variance'' of approximation error}
\label{appendix:variance}
In this section, to justify the use of the average effect derived in Sec.~\ref{sec:average_SK}, we compute ``variance'' of this random effect. 
In this subsection, let $\mathbb{E}$ denote the average within the constraint Eq.~\eqref{eq:Rn_approx}. Now, we compute a ``variance'' operator 
\begin{equation*}
    \mathbb{E} [(R_{\bm{n}}(\varphi) AR_{\bm{n}}^\dagger (\varphi) -  \mathbb{E}[R_{\bm{n}}(\varphi) A R_{\bm{n}}^\dagger (\varphi)])(R_{\bm{n}}(\varphi) A R_{\bm{n}}^\dagger (\varphi) -  \mathbb{E}[R_{\bm{n}}(\varphi) A R_{\bm{n}}^\dagger (\varphi)])^\dagger]
\end{equation*}
Since 
\begin{equation*}
    \mathbb{E}[R_{\bm{n}}(\varphi) A R_{\bm{n}}^\dagger (\varphi)] = (1-p)A + p\tr(A)\frac{I}{2}
\end{equation*}
as we derived above, the ``variance'' operator is 
\begin{align*}
    &\mathbb{E} \left[\left(R_{\bm{n}}(\varphi) AR_{\bm{n}}^\dagger (\varphi) -  (1-p)A + p\tr(A)\frac{I}{2}\right)^\dagger\left(R_{\bm{n}}(\varphi) A R_{\bm{n}}^\dagger (\varphi) - (1-p)A + p\tr(A)\frac{I}{2}\right)\right] \\ 
    &=\mathbb{E} \left[R_{\bm{n}}(\varphi) A^\dagger A R_{\bm{n}}^\dagger (\varphi)\right] - \left((1-p)A + p\tr(A)\frac{I}{2}\right)^\dagger \left((1-p)A + p\tr(A)\frac{I}{2}\right) \\ 
    &= (1-p)A^\dagger A + p\tr(A^\dagger A)\frac{I}{2}\\ 
    &\quad- \left((1-p)^2A^\dagger A+  \frac{p(1-p)}{2}(\tr(A^\dagger)A + \tr(A)A^\dagger) +p^2|\tr(A)|^2 \frac{I}{4}\right) \\ 
    &=p\left((1-p)A^\dagger A - \frac{(1-p)}{2}(\tr(A^\dagger)A + \tr(A)A^\dagger) +\left(\tr(A^\dagger A) - \frac{p|\tr(A)|^2}{2} \right) \frac{I}{2}\right). 
\end{align*}
The trace norm of this operator is 
\begin{equation*}
    p\left((2-p)\tr(A^\dagger A) + \left(\frac{p}{2} - 1\right)|\tr(A)|^2 \right) = O(p) = O(\epsilon^2).  
\end{equation*}
In this sense, the variance of the approximation error is small when each gate-approximation error $\epsilon$ is sufficiently small. 

\newpage 
\section{Unitary coupled-cluster singles and doubles ansatz}~\label{sec:AppendixA}
\newcommand{\bmth}{\bm{\theta}}

In the numerical simulations in the main text, we employed unitary coupled-cluster singles and doubles (UCCSD) ansatz~\cite{Anand2022}.
The UCCSD ansatz is defined by 
\begin{align}
 \hat{U}_\mathrm{UCCSD}(\bmth) &= \mathrm{e}^{\hat{T}_s - \hat{T}_s^\dag} \mathrm{e}^{\hat{T}_{d1} - \hat{T}_{d1}^\dag} \mathrm{e}^{\hat{T}_{d2} - \hat{T}_{d2}^\dag}, \\
 \hat{T}_s &= \sum_{o,v} \theta_{ov}^{(s)} \left(  \hat{a}_{v\uparrow}^\dag \hat{a}_{o\uparrow} + \hat{a}_{v\downarrow}^\dag \hat{a}_{o\downarrow} \right),\\
 \hat{T}_{d1} &= \sum_{o, v} \theta^{(d1)}_{ov} \hat{a}_{v\uparrow}^\dag \hat{a}_{o\uparrow} \hat{a}_{v\downarrow}^\dag \hat{a}_{o\downarrow}, \\
 \hat{T}_{d2} &= \sum_{(o_1, v_1) \neq (o_2, v_2)} \sum_{\sigma, \tau = \uparrow, \downarrow} \theta_{o_1 o_2 v_1 v_2}^{(d2)} \hat{a}_{v_1\sigma}^\dagger \hat{a}_{o_1\sigma} \hat{a}_{v_2\tau}^\dagger \hat{a}_{o_2\tau},
\end{align}
where $a_{p\sigma} (a_{p\sigma}^\dag)$ is an annihilation (creation) operator of electron in an orbital $p$ with spin $\sigma$ and $o,o_1,o_2$ ($v,v_1,v_2$) are the occupied (virtual) orbitals in the Hartree-Fock state.
The circuit parameters of this ansatz are $\bm{\theta} = (\theta_{ov}^{(s)})_{ov} \cup (\theta_{ov}^{(d1)})_{ov} \cup (\theta_{o_1o_2v_1v_2}^{(d2)})_{o_1o_2v_1v_2}$.
To transform this ansatz into qubit operators, we performed the Jordan-Wigner transformation~\cite{Jordan1928} for $\hat{T}_s - \hat{T}_s^\dag, \hat{T}_{d1} - \hat{T}_{d1}^\dag$ and $\hat{T}_{d2} - \hat{T}_{d2}^\dag$.
For example, if $\hat{T}_s - \hat{T}_s^\dag$ is transformed into the sum of multiqubit Pauli operators $i\sum_{ov} \theta_{ov}^{(s)} \sum_k \hat{P}_k^{(ov)}$, we (approximately) implement $e^{\hat{T}_s - \hat{T}_s^\dag}$ by the Trotterization:
\begin{equation*}
 \prod_{ov} \prod_k \exp[i\theta_{ov}^{(s)} \hat{P}_k^{(ov)}].
\end{equation*}
The operators $e^{\hat{T}_{d1} - \hat{T}_{d1}^\dag}$ and $e^{\hat{T}_{d2} - \hat{T}_{d2}^\dag}$ were implemented as qubit operators in the same way.

Note that every multi-qubit Pauli rotation $\exp[i\theta_{ov}^{(s)} \hat{P}_k^{(ov)}]$ can be expressed as a composite of Clifford gates and a single-qubit $z$-rotation with the same angle $\theta_{ov}$~\cite{Anand2022}. 
Therefore, it suffices to give the Clifford+$T$ decomposition of the corresponding $z$-rotation to obtain the decomposition of $\exp[i\theta_{ov}^{(s)} \hat{P}_k^{(ov)}]$. 

\newpage
\section{Results of Numerical Experiments}~\label{sec:AppendixB}
In this section, we show the whole results of our numerical experiments, which are partly shown in Sec.~\ref{subsec:numerical_experiments}. 
We consider the molecules shown in Table~\ref{tab: geometries}. 

\begin{figure}[htbp]
    \begin{tabular}{ccc}
      \begin{subfigure}[t]{0.33\textwidth}
        \centering
        \includegraphics[keepaspectratio, width=\linewidth]{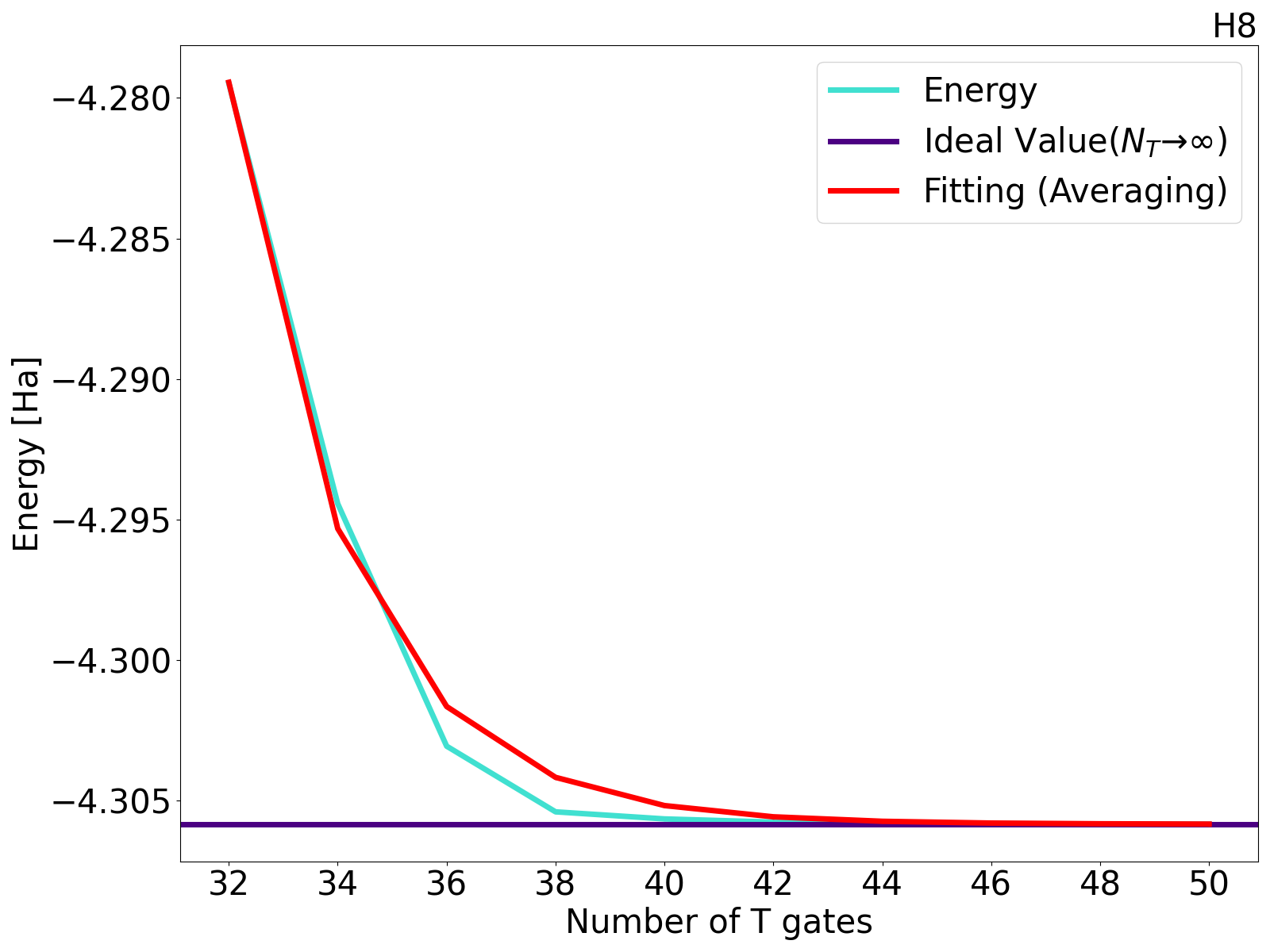}
        \subcaption{\ce{H_8}. The fitting parameter is $c = 51(2)$.}
      \end{subfigure} &
    \begin{subfigure}[t]{0.33\textwidth}
        \centering
        \includegraphics[keepaspectratio, width=\linewidth]{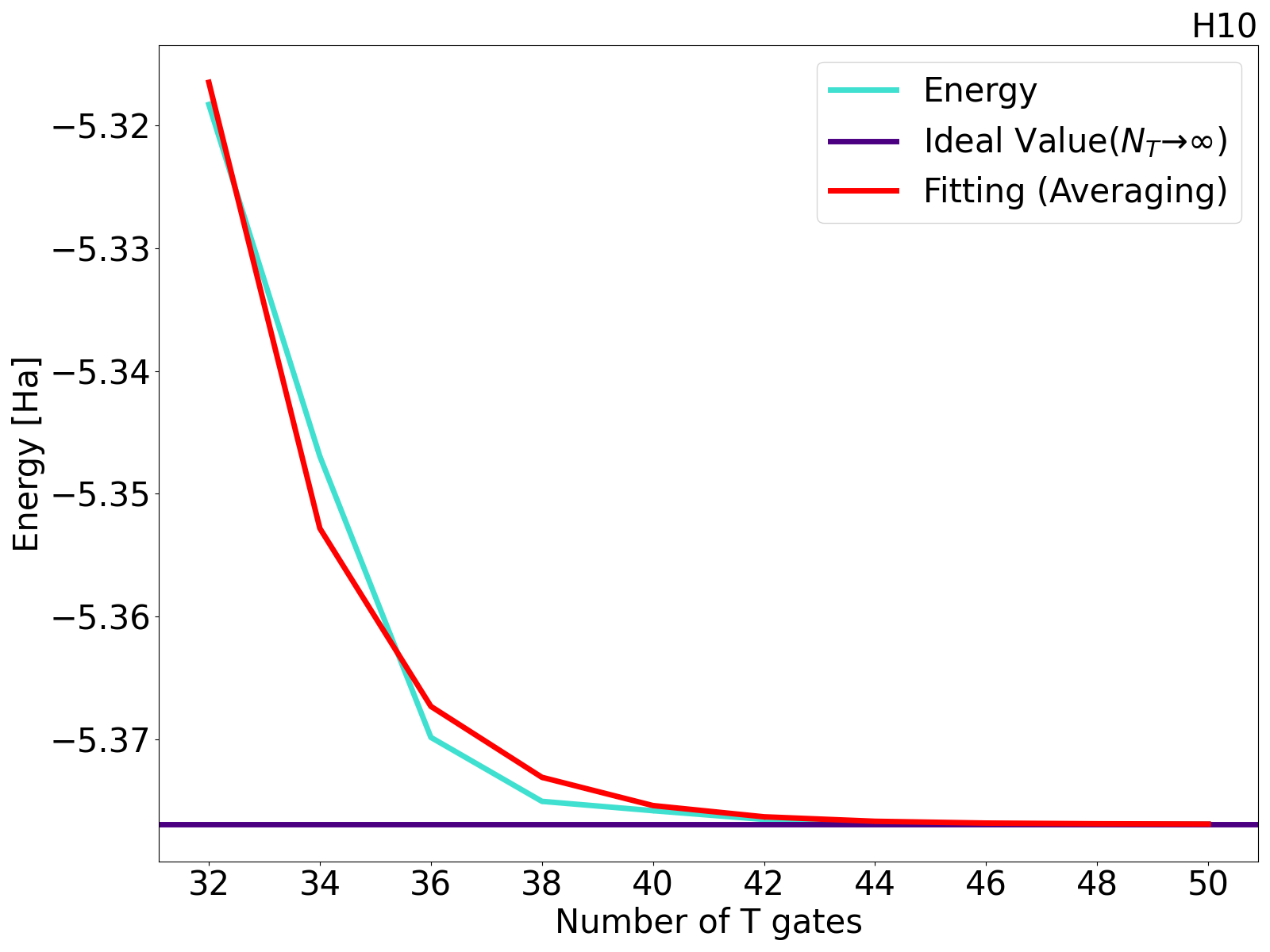}
        \subcaption{\ce{H_{10}}. The fitting parameter is $c = 40(2)$.}
      \end{subfigure} &
      \begin{subfigure}[t]{0.33\textwidth}
        \centering
        \includegraphics[keepaspectratio, width=\linewidth]{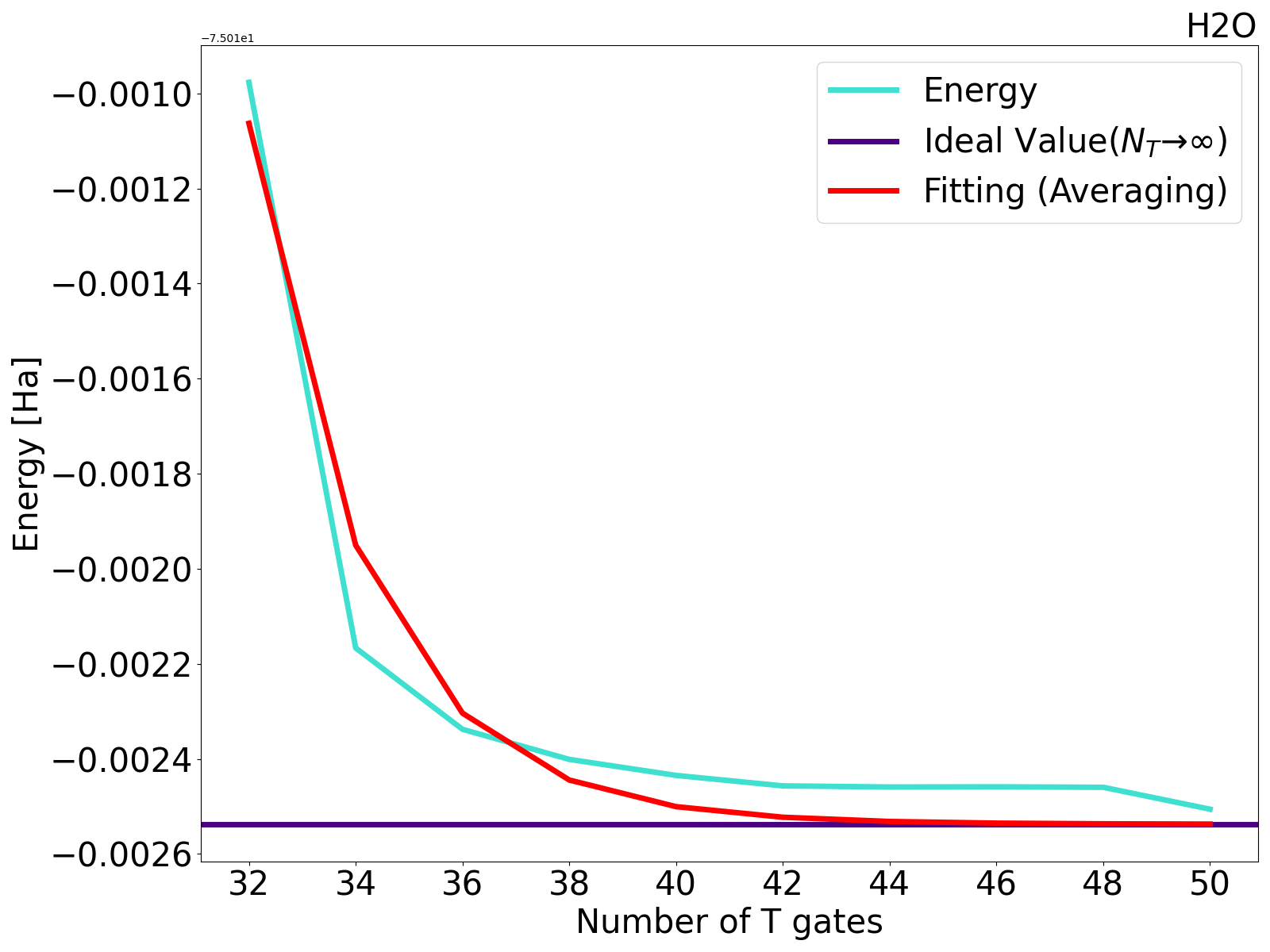}
        \subcaption{\ce{H_2O}. The fitting parameter is $c = 0.3799(5)$.}
      \end{subfigure} \\
   
      \begin{subfigure}[t]{0.33\textwidth}
        \centering
        \includegraphics[keepaspectratio, width=\linewidth]{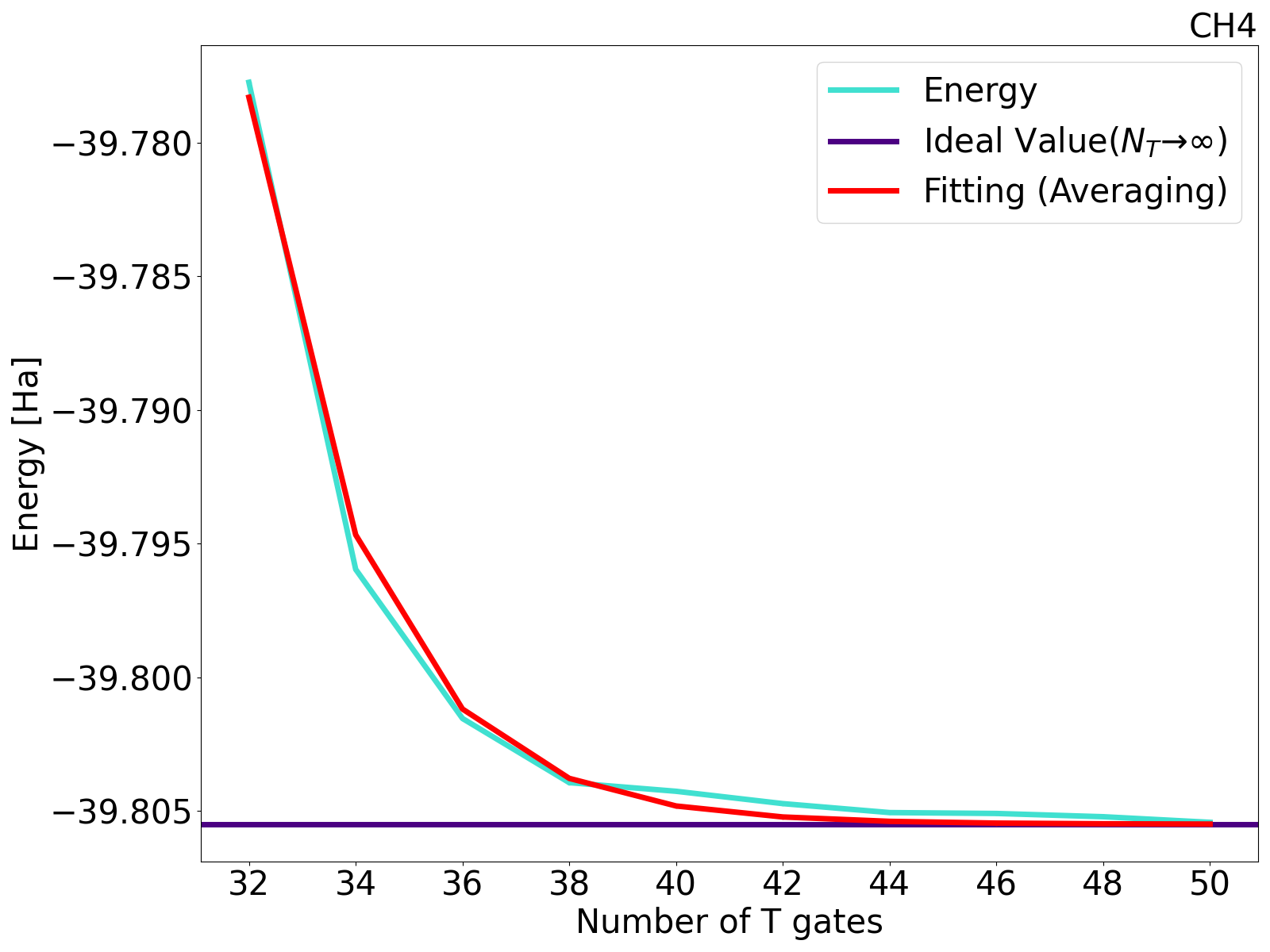}
        \subcaption{\ce{CH_4}. The fitting parameter is $c = 3.730(5)$.}
      \end{subfigure} &
    \begin{subfigure}[t]{0.33\textwidth}
        \centering
        \includegraphics[keepaspectratio, width=\linewidth]{figs/C2H2_SK_energy_ccsd.png}
        \subcaption{\ce{C_2H_2}. The fitting parameter is $c = 0.6106(2)$.}
      \end{subfigure} &
      \begin{subfigure}[t]{0.33\textwidth}
        \centering
        \includegraphics[keepaspectratio, width=\linewidth]{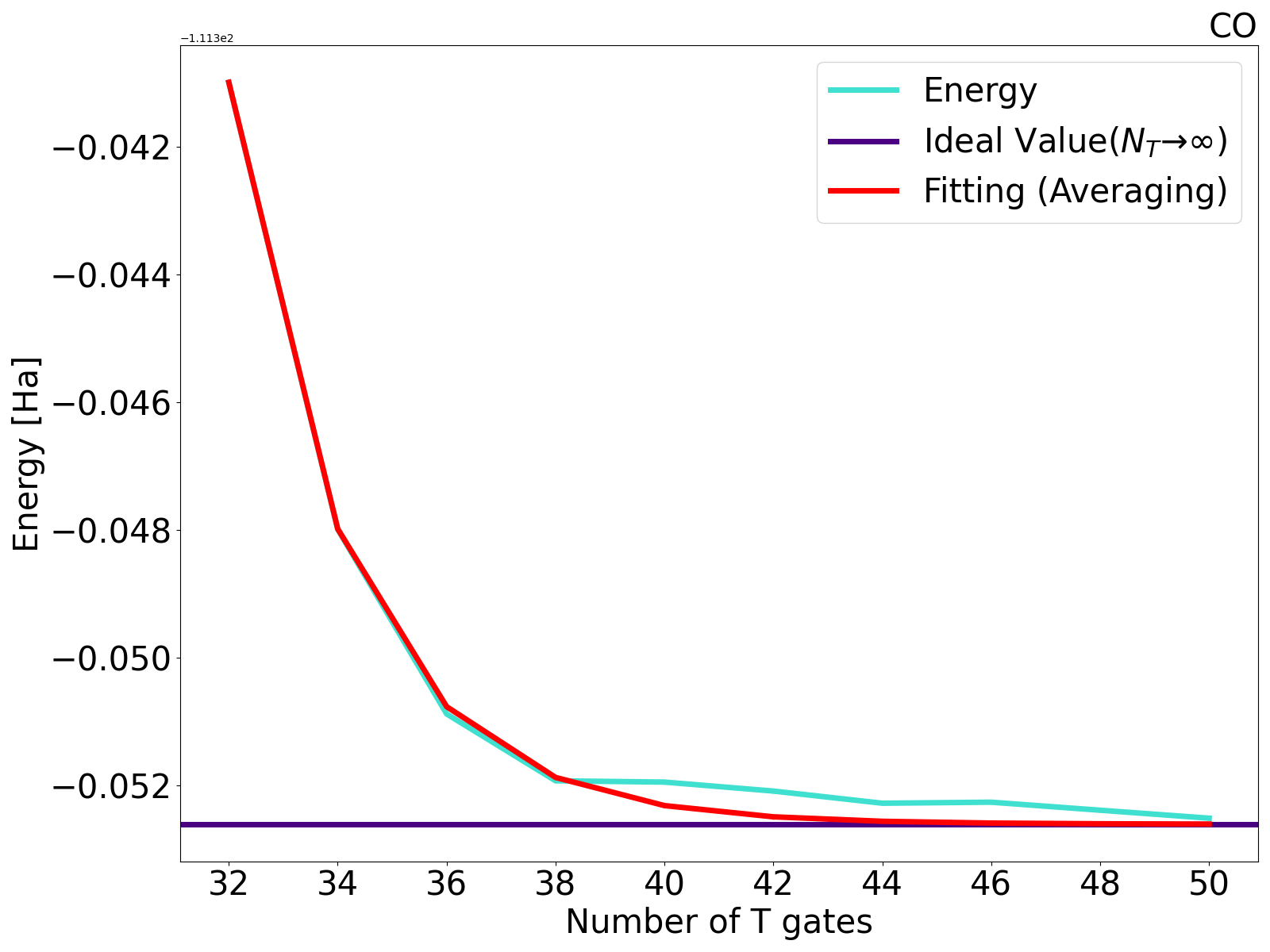}
        \subcaption{\ce{CO}. The fitting parameter is $c = 0.5196(1)$.}
      \end{subfigure} \\
      
            \begin{subfigure}[t]{0.33\textwidth}
        \centering
        \includegraphics[keepaspectratio, width=\linewidth]{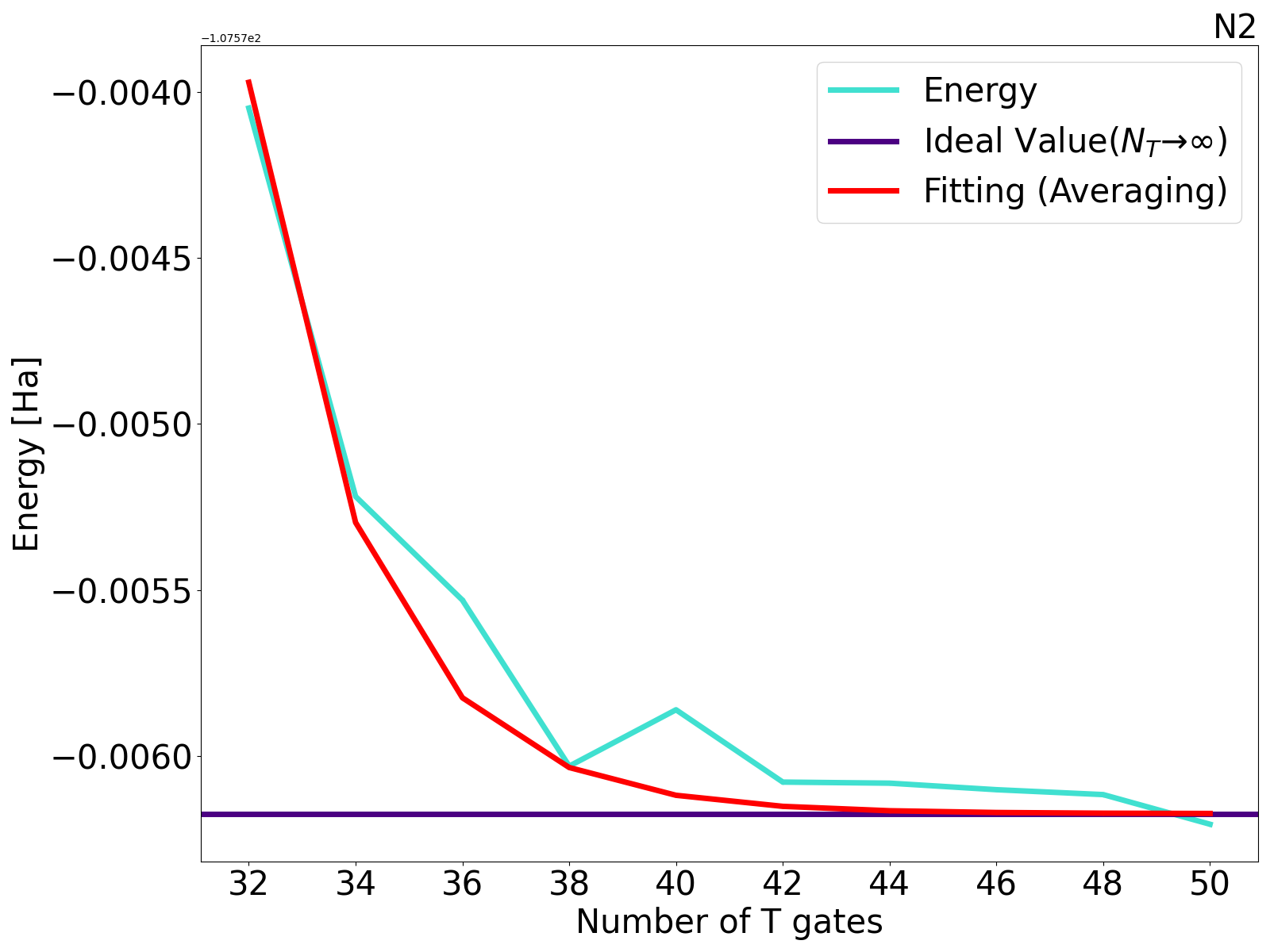}
        \subcaption{\ce{N_2}. The fitting parameter is $c = 0.10216(3)$.}
      \end{subfigure} &
    \begin{subfigure}[t]{0.33\textwidth}
        \centering
        \includegraphics[keepaspectratio, width=\linewidth]{figs/NH3_SK_energy_ccsd.png}
        \subcaption{\ce{NH_3}. The fitting parameter is $c = 0.779(2)$.}
      \end{subfigure} & 
      \\
      \end{tabular}
     \caption{Energy expectation values of Clifford+$T$-gate decomposed UCCSD' states with $T$-gate budget $32 \leq N_T \leq 50$. In each panel, the orange curve represents the expectation values of the UCCSD' state that are decomposed by Algorithm~\ref{alg:ross}. The horizontal purple line indicates the true UCCSD' state energy obtained by the UCCSD ansatz without Clifford+$T$ decomposition. The blue curve is the fitting of the orange curve modeled by Eq.~\eqref{eq:model}. }
        \label{fig:energy_app}
  \end{figure}

In Fig.~\ref{fig:energy_app}, we show the energy curve of each molecule, and in Fig.~\ref{fig:fidelity_app} we show the fidelity curve and indicate the threshold $T$-count to achieve the relative error $\delta_{\textup{rel}} < 10^{-4}$. 

\begin{figure}[htbp]
    \begin{tabular}{ccc}
      \begin{subfigure}[t]{0.33\textwidth}
        \centering
        \includegraphics[keepaspectratio, width=\linewidth]{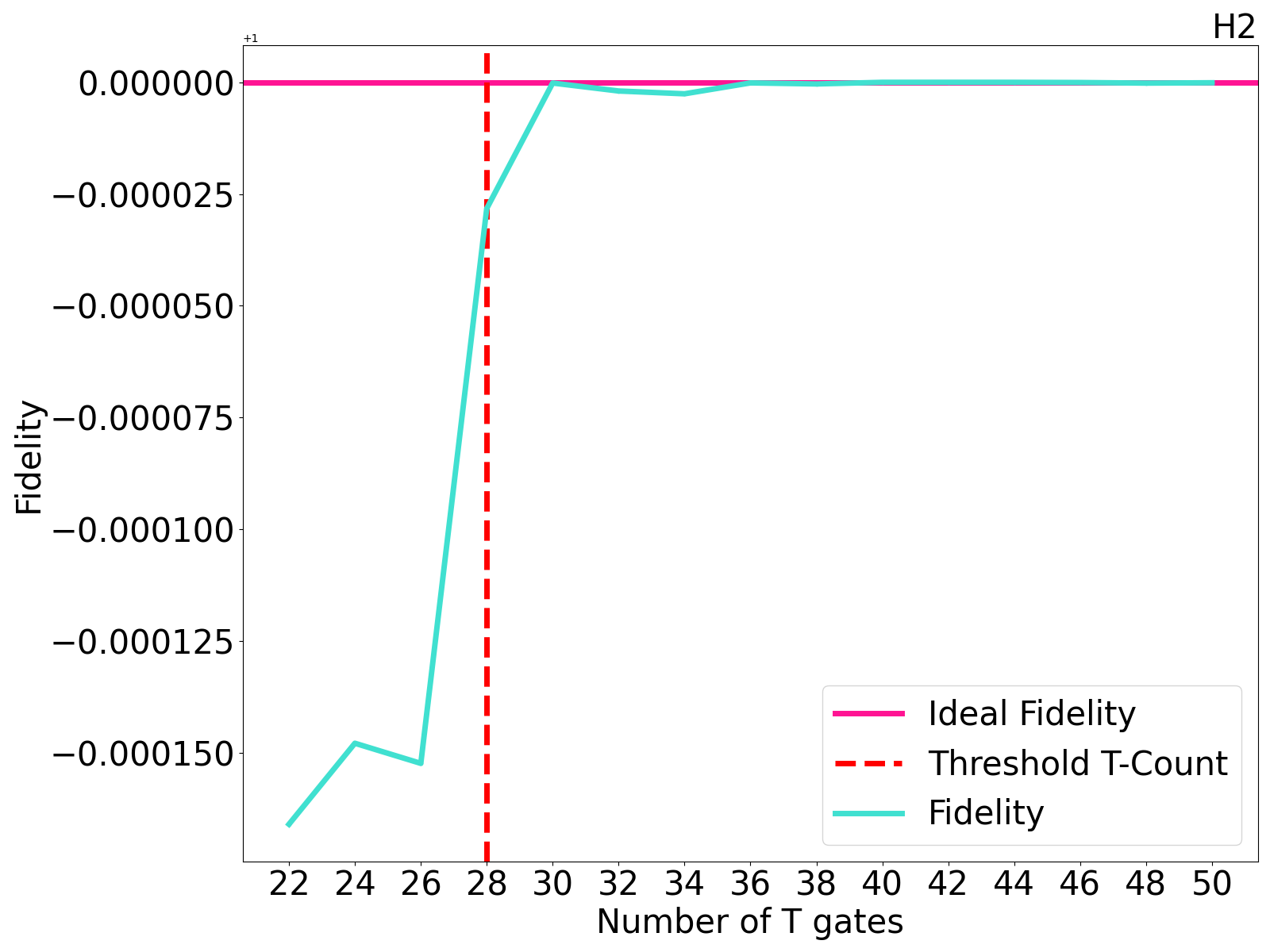}
        \subcaption{\ce{H_2}}
      \end{subfigure} &
    \begin{subfigure}[t]{0.33\textwidth}
        \centering
        \includegraphics[keepaspectratio, width=\linewidth]{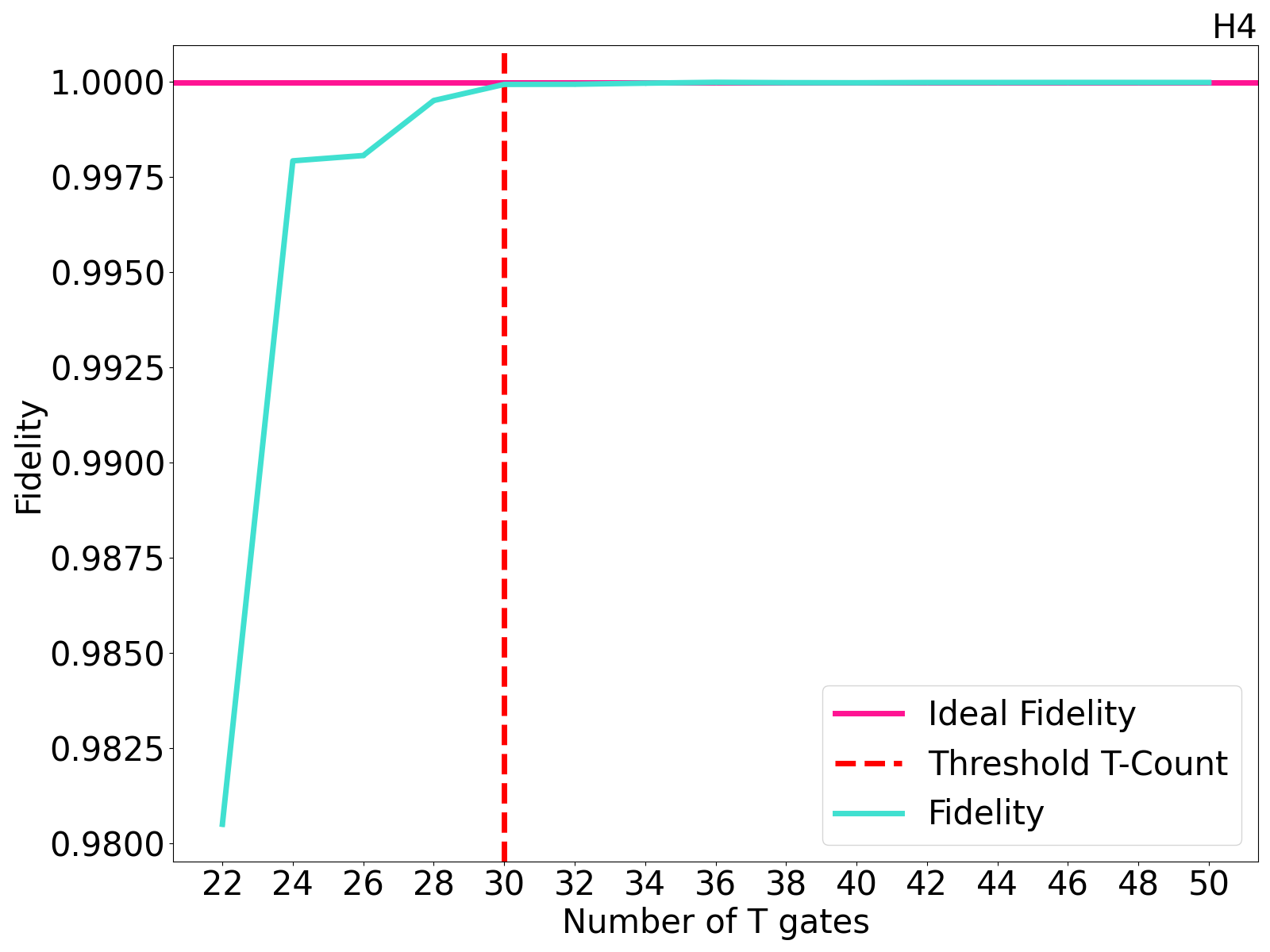}
        \subcaption{\ce{H_4}}
      \end{subfigure} &
      \begin{subfigure}[t]{0.33\textwidth}
        \centering
        \includegraphics[keepaspectratio, width=\linewidth]{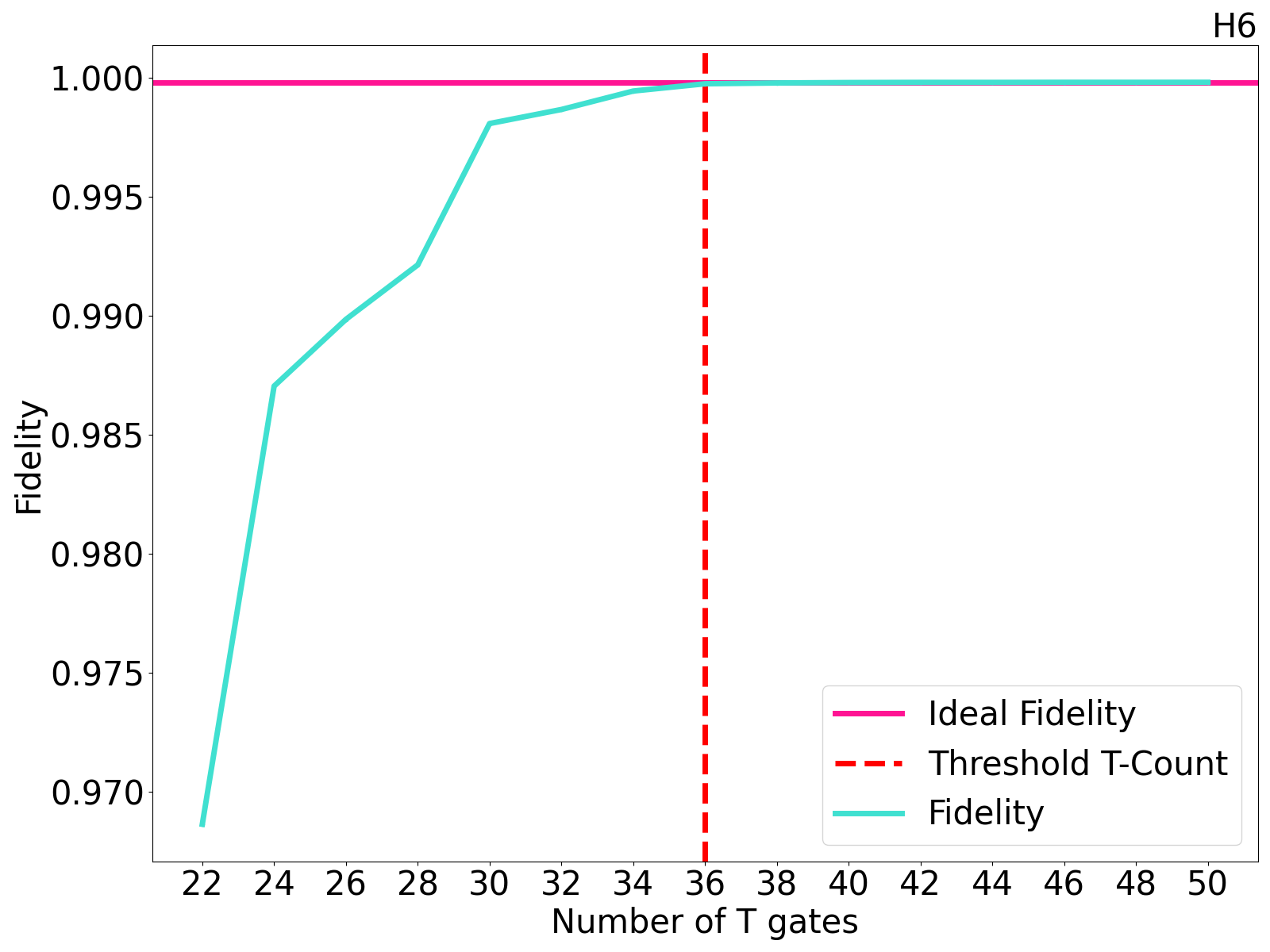}
        \subcaption{\ce{H_6}}
      \end{subfigure} \\
   
      \begin{subfigure}[t]{0.33\textwidth}
        \centering
        \includegraphics[keepaspectratio, width=\linewidth]{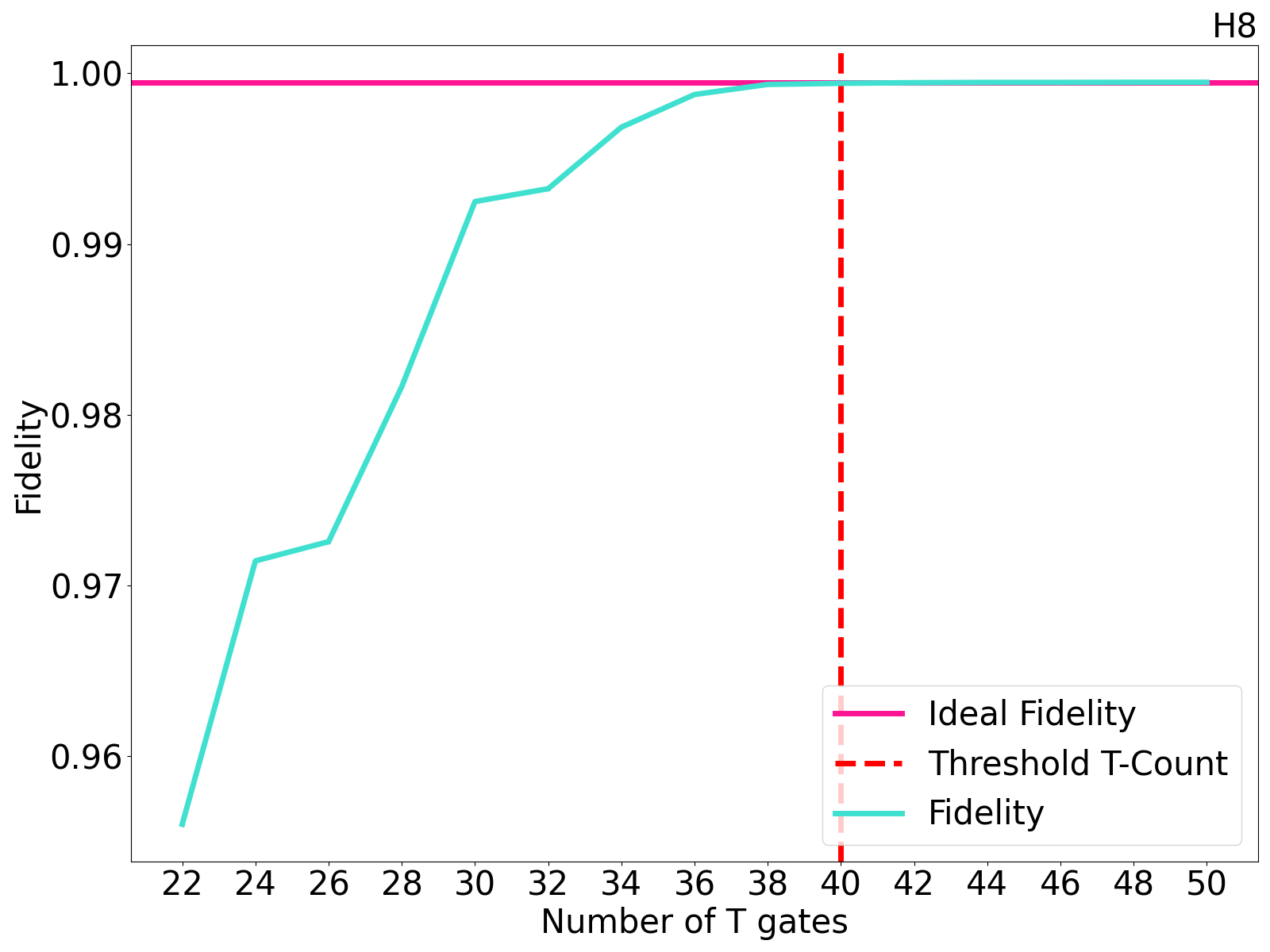}
        \subcaption{\ce{H_8}}
      \end{subfigure} &
      \begin{subfigure}[t]{0.33\textwidth}
        \centering
        \includegraphics[keepaspectratio, width=\linewidth]{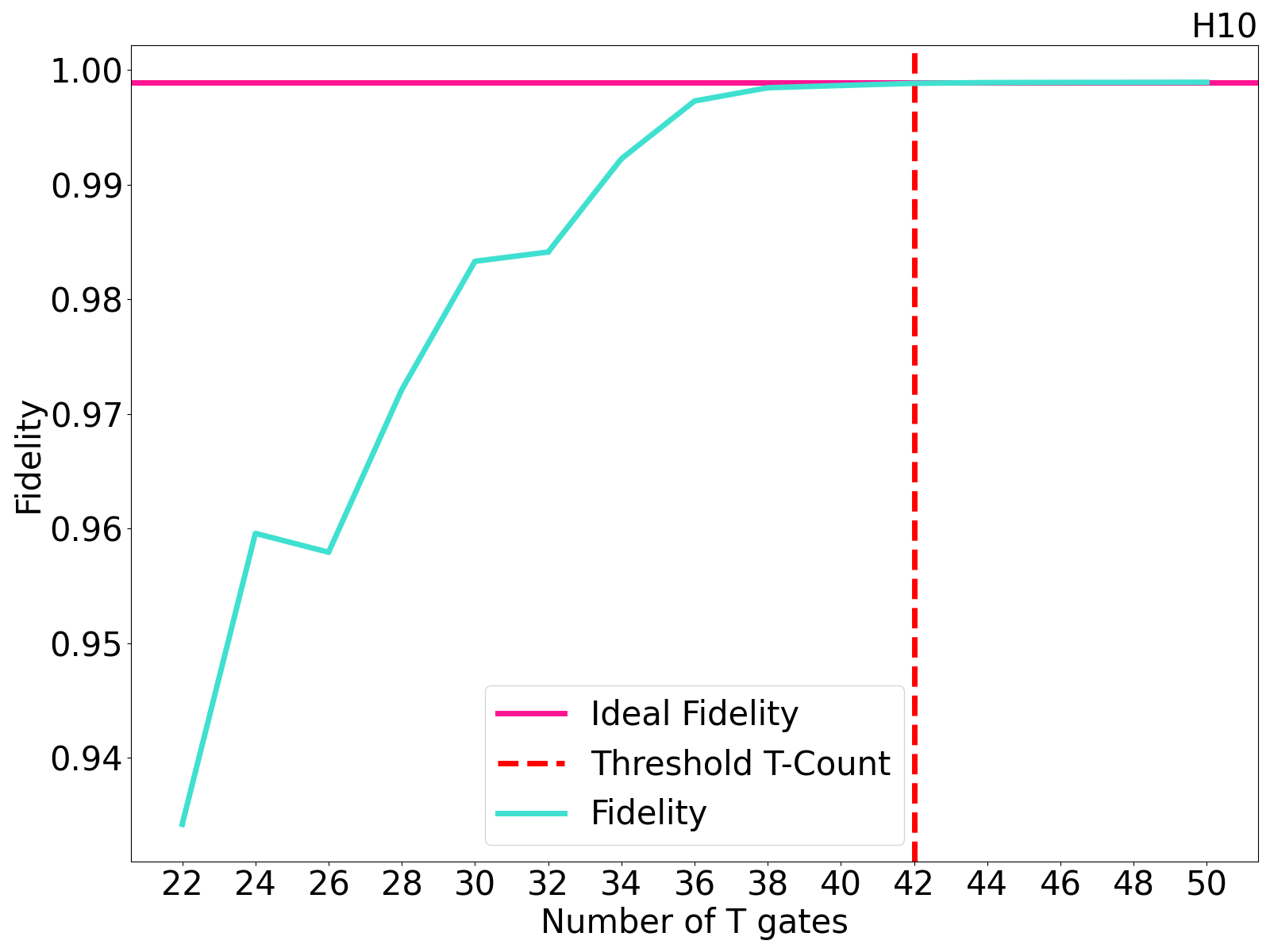}
        \subcaption{\ce{H_{10}}}
      \end{subfigure} &
      \begin{subfigure}[t]{0.33\textwidth}
        \centering
        \includegraphics[keepaspectratio, width=\linewidth]{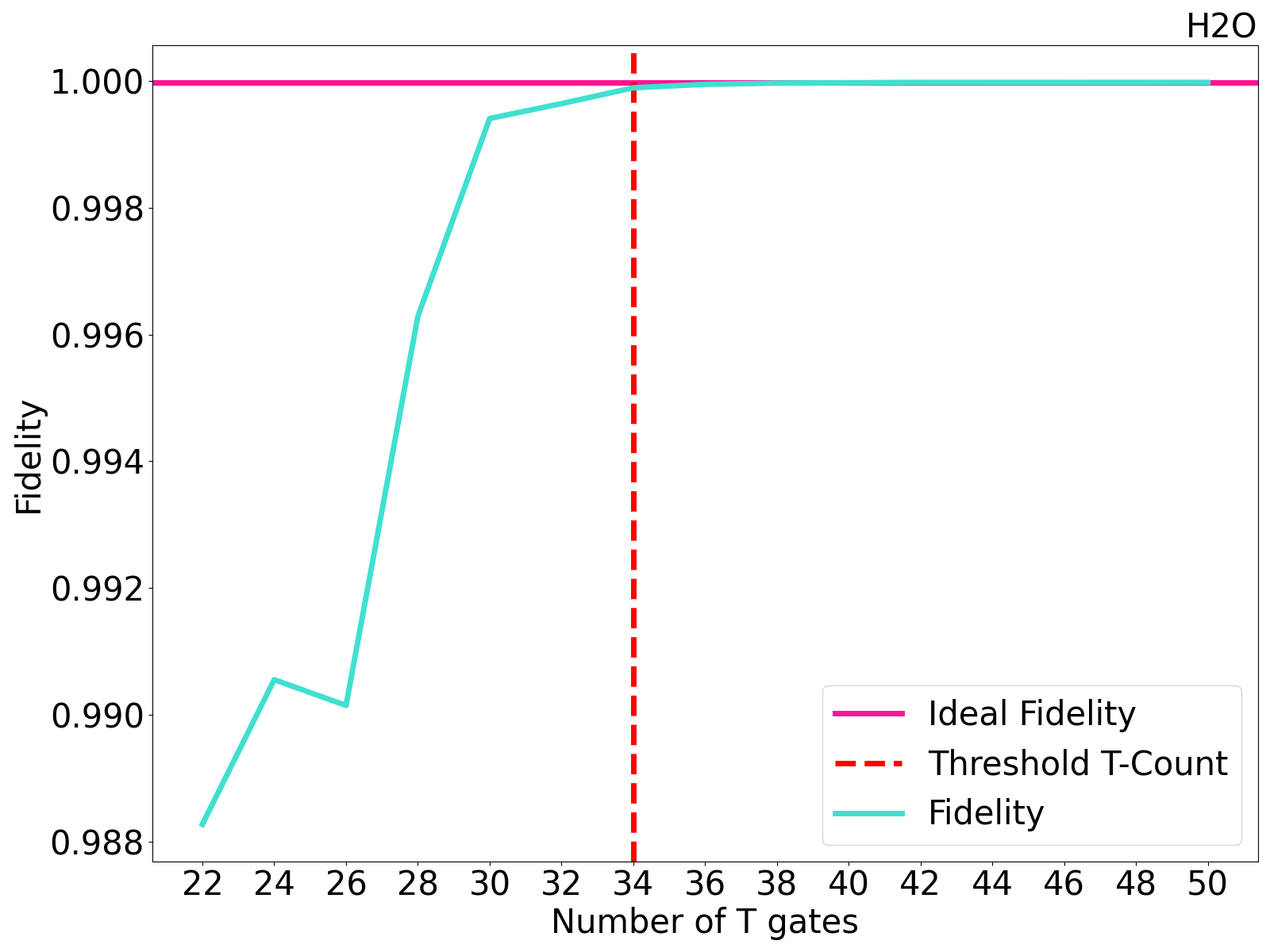}
        \subcaption{\ce{H_2O}}
      \end{subfigure} \\
      
    \begin{subfigure}[t]{0.33\textwidth}
        \centering
        \includegraphics[keepaspectratio, width=\linewidth]{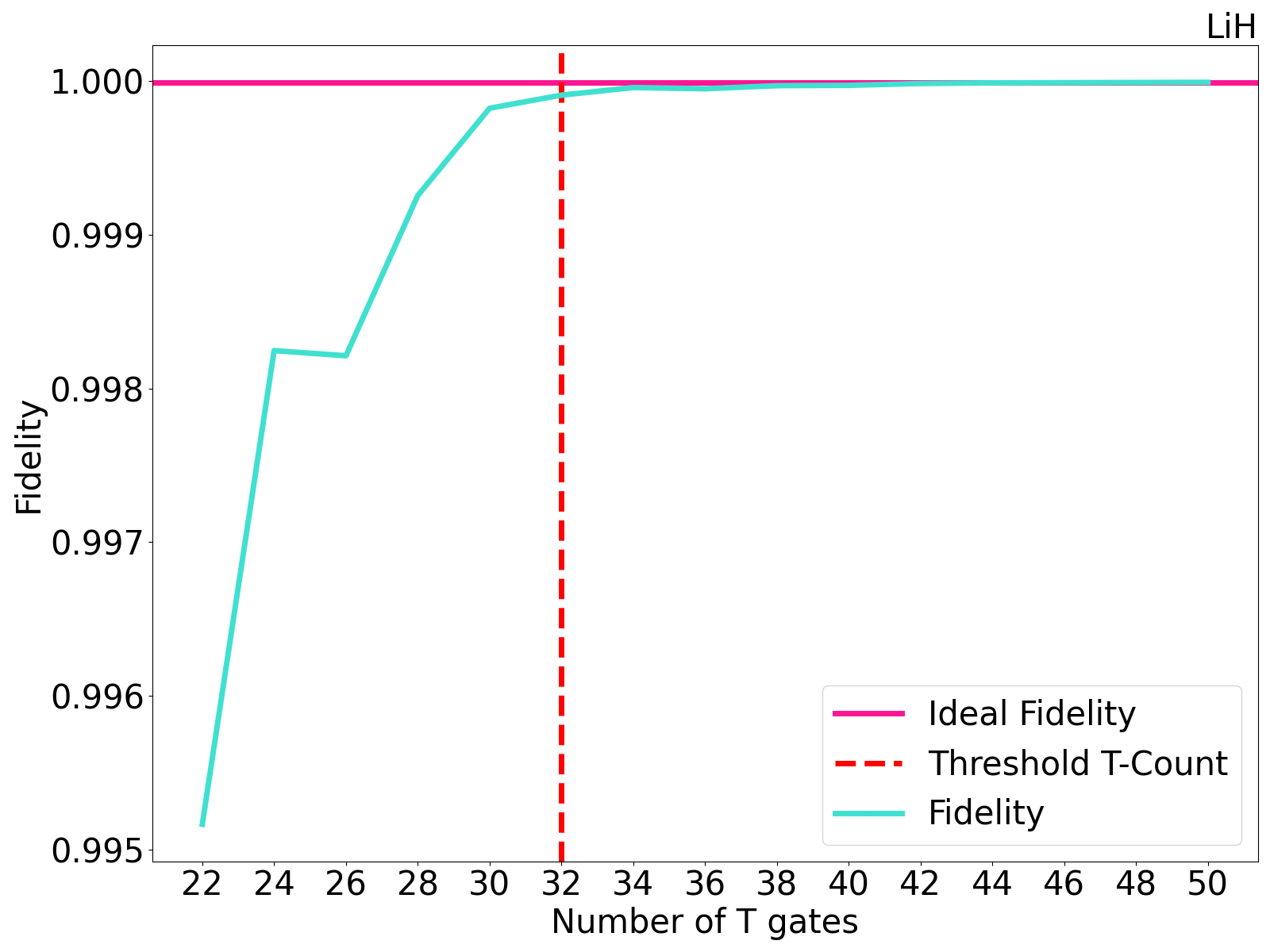}
        \subcaption{\ce{LiH}}
      \end{subfigure} &
      \begin{subfigure}[t]{0.33\textwidth}
        \centering
        \includegraphics[keepaspectratio, width=\linewidth]{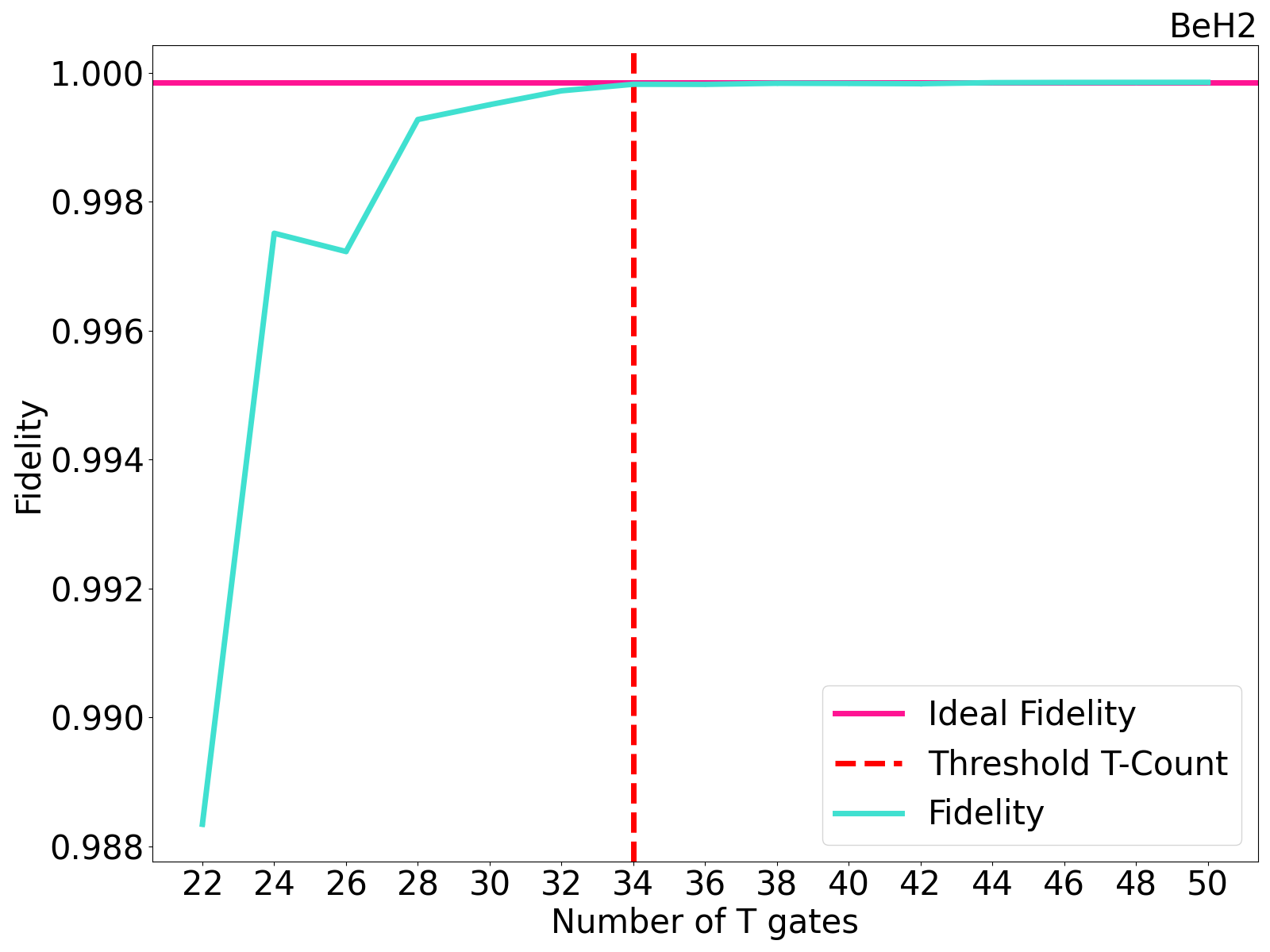}
        \subcaption{\ce{BeH_{2}}}
      \end{subfigure} &
      \begin{subfigure}[t]{0.33\textwidth}
        \centering
        \includegraphics[keepaspectratio, width=\linewidth]{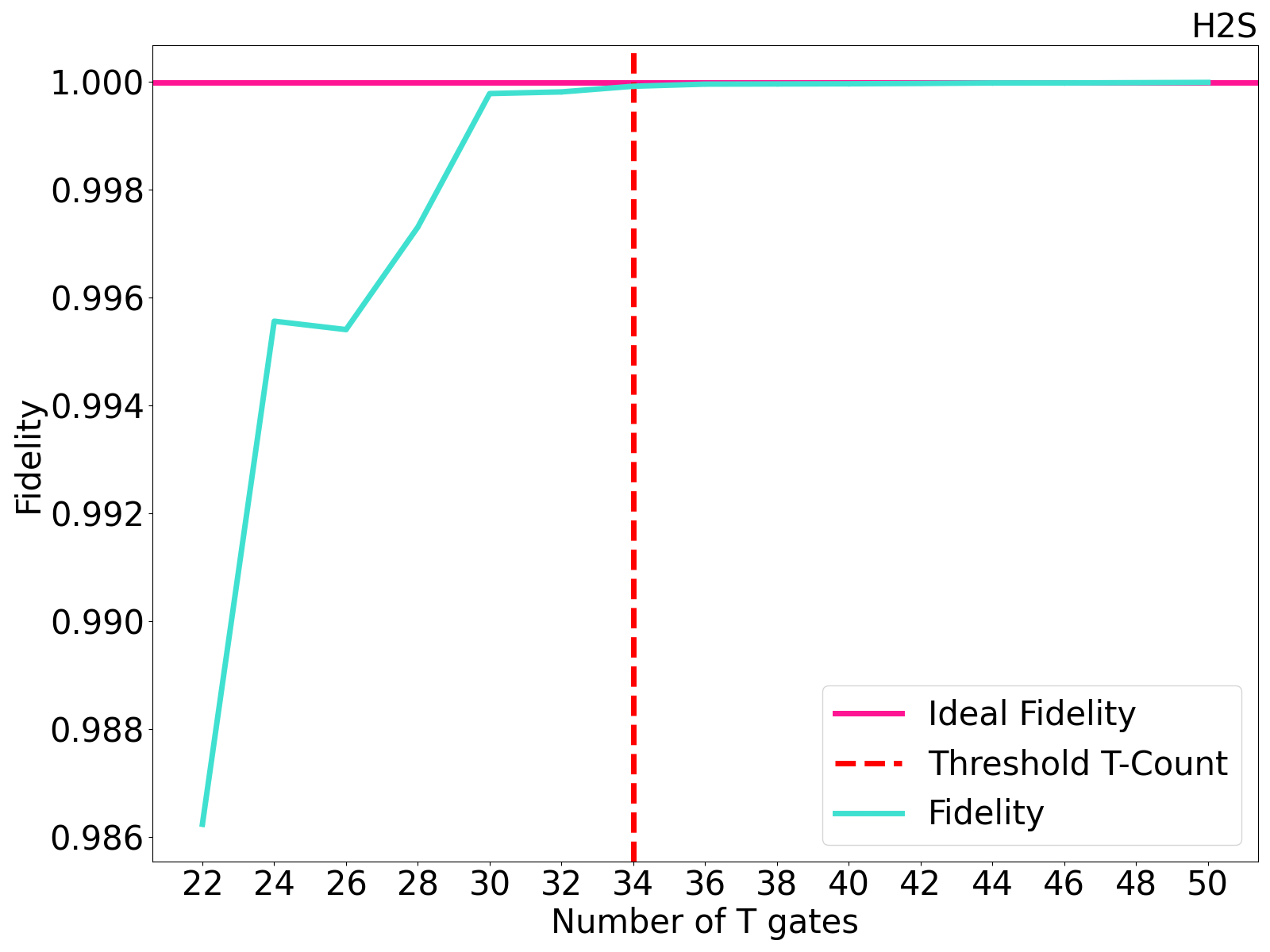}
        \subcaption{\ce{H_2S}}
      \end{subfigure} \\
      
        \begin{subfigure}[t]{0.33\textwidth}
        \centering
        \includegraphics[keepaspectratio, width=\linewidth]{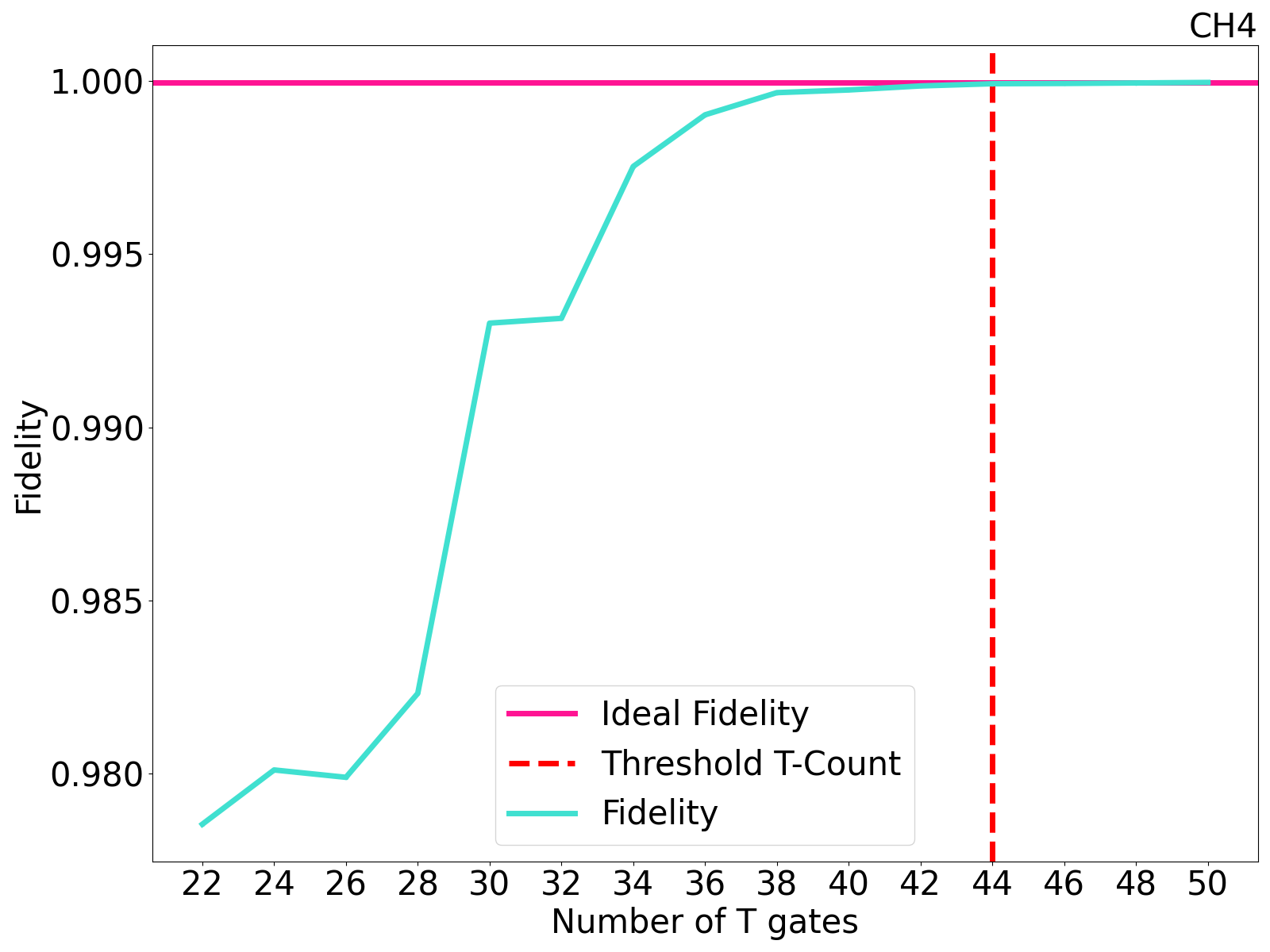}
        \subcaption{\ce{CH_4}}
      \end{subfigure} &
      \begin{subfigure}[t]{0.33\textwidth}
        \centering
        \includegraphics[keepaspectratio, width=\linewidth]{figs/C2H2_SK_fidelity_ccsd.png}
        \subcaption{\ce{C_2H_{2}}}
      \end{subfigure} &
      \begin{subfigure}[t]{0.33\textwidth}
        \centering
        \includegraphics[keepaspectratio, width=\linewidth]{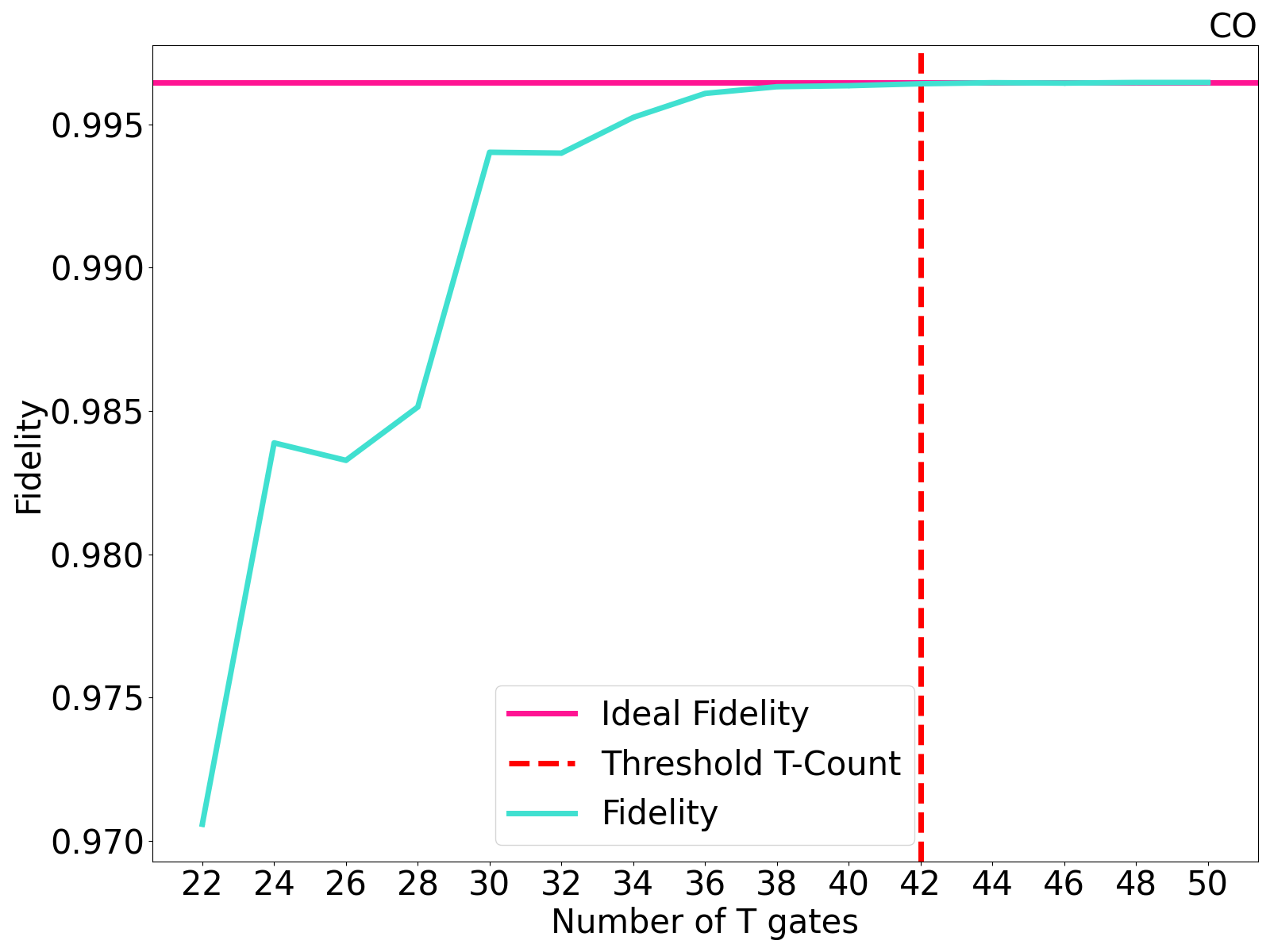}
        \subcaption{\ce{CO}}
      \end{subfigure} 
    \end{tabular}
\end{figure}

\setcounter{figure}{4}
\begin{figure}[H]
\begin{tabular}{ccc}
    \begin{subfigure}[b]{0.33\textwidth}
     \addtocounter{subfigure}{12}
        \centering
        \includegraphics[keepaspectratio, width=\linewidth]{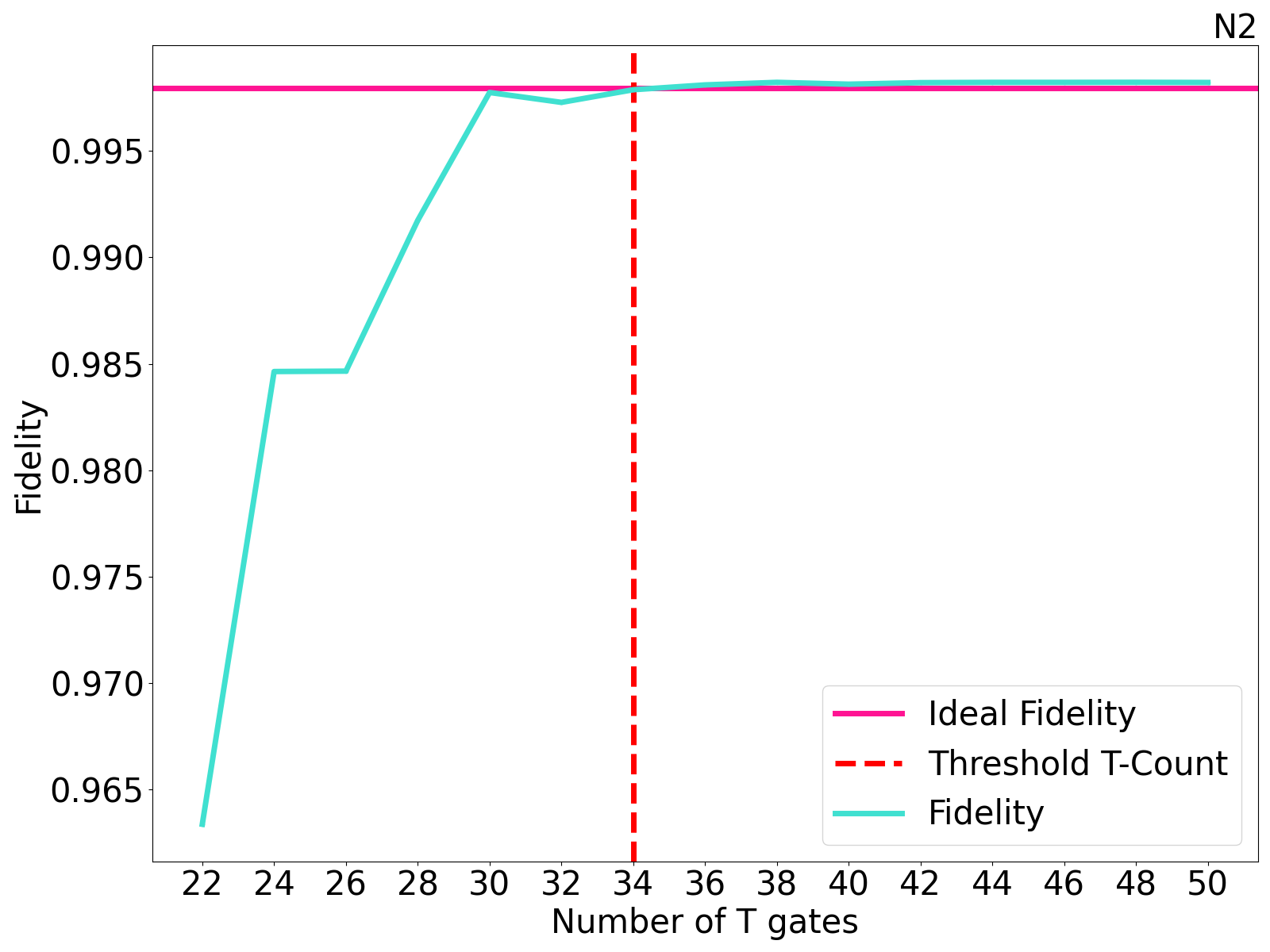}
        \subcaption{\ce{N_2}}
      \end{subfigure} &
      \begin{subfigure}[b]{0.33\textwidth}
        \centering
        \includegraphics[keepaspectratio, width=\linewidth]{figs/NH3_SK_fidelity_ccsd.png}
        \subcaption{\ce{NH_3}}
      \end{subfigure} & 
    \end{tabular}
     \caption{The fidelity between the exact ground state of the Hamiltonian and the  Clifford+$T$-gate decomposed UCCSD' states with $T$-gate budget $22 \leq N_T \leq 50$. In each panel, the blue curve represents the fidelity between the exact ground state and the UCCSD' state that is decomposed by Algorithm~\ref{alg:ross}. The horizontal pink line indicates the fidelity between the exact ground state and the true UCCSD' state obtained by the UCCSD ansatz without Clifford+$T$ decomposition. The orange vertical line represents the $T$-count threshold that achieves Eq.~\eqref{eq:rel_error}.}
        \label{fig:fidelity_app}
  \end{figure}
\end{document}